\documentclass[10pt,twocolumn,aps,pra,groupedaddress,amsmath,amssymb,english]{revtex4-2}
\usepackage{graphicx}
\usepackage{dcolumn}
\usepackage{bm}
\usepackage[T1]{fontenc}
\usepackage{textcomp}
\usepackage{mathrsfs}
\usepackage{amsmath}
\usepackage{bbm}
\usepackage[normalem]{ulem}
\usepackage{amsbsy}
\usepackage{braket}
\usepackage{amstext}
\usepackage{amssymb}
\usepackage{setspace}
\usepackage{esint}
\usepackage{xcolor,colortbl}
\usepackage[colorlinks,citecolor=blue,urlcolor=blue,linkcolor=blue,hypertexnames=true]{hyperref} 
\usepackage{babel}
\usepackage{booktabs}
\usepackage{mciteplus}
\begin{document}

\title{Entangling capability of coherently controlled quantum processes}

\author{Beyza Aslanbaş}
\affiliation{Faculty of Engineering and Natural Sciences, Sabanci University, Tuzla, Istanbul 34956, Turkey}

\author{Bedirhan Alkan}
\affiliation{Faculty of Engineering and Natural Sciences, Sabanci University, Tuzla, Istanbul 34956, Turkey}

\author{G\"{o}ktu\u{g} Karpat}
\email{goktug.karpat@sabanciuniv.edu}
\affiliation{Faculty of Engineering and Natural Sciences, Sabanci University, Tuzla, Istanbul 34956, Turkey}

\date{\today}

\begin{abstract}
We analytically characterize entanglement generation by two paradigmatic coherently controlled quantum processes, the quantum switch and time-flip. Retaining rather than measuring or discarding the control, we treat the control and target as the bipartite system and assume pure product inputs, so that any output entanglement reflects the entangling capability of the process. For the switch of qubit unitaries, we derive exact expressions for entanglement, together with a geometric characterization and a coherence–entanglement conservation relation. We then extend our analysis to binary random unitary, Pauli, amplitude damping, and generalized amplitude damping channels. We prove that switch of channels commuting under composition cannot entangle a separable input. Yet maximal entanglement is possible even with dissipation, e.g., an entanglement-breaking damping channel switched with a bit-flip can transform a product input into a maximally entangled state. For two generalized amplitude damping channels, a common stationary state prevents entanglement, while different stationary populations can enable it. We  present an analogous study for the time-flip of unitary and binary random unitary channels. Our findings provide a systematic analysis of how much entanglement can be generated by coherent control of channel order or input-output direction.

\end{abstract}

\maketitle

\section{Introduction}

Quantum theory permits coherent control not only of which operations act on a system, but also of the order in which they can act. Higher-order quantum maps provide the general framework for such transformations~\cite{Chiribella2008}, and their paradigmatic example is the
quantum switch, where a control system coherently determines the order in which two quantum operations
act on a target~\cite{Chiribella2013,Dong2023}. The quantum switch and related controlled processes~\cite{Oreshkov2012,Araujo2015} have been investigated across a broad range of settings, including channel discrimination~\cite{Chiribella2012,Bavaresco2021}, query and computation complexity~\cite{Colnaghi2012,Araujo2014,Araujo2017,Taddei2021,Renner2022,Simonov2026}, communication complexity~\cite{Guerin2016}, communication over noisy quantum channels~\cite{Ebler2018,Guerin2019,Procopio2019,Abbott2020,Goswami2020a,Caleffi2020,Mukhopadhyay2020,Bhattacharya2021,Chiribella2021a,Sazim2021,Chiribella2021b,Wu2025,Subesh2026}, quantum thermodynamics~\cite{Felce2020,Guha2020,Liu2022,Francica2022,Simonov2022,Dieguez2023,Goldberg2023,Zhu2023,Molitor2024,Simonov2025,Sharma2026,Lisboa2026}, non-Markovian quantum processes~\cite{Maity2024,Mukherjee2024,Karpat2024,Anand2025} and quantum metrology~\cite{Zhao2020,Chapeau-Blondeau2021,Liu2023}. It has been implemented in photonic architectures~\cite{Procopio2015,Rubino2017,Goswami2018,Wei2019,Guo2020,Cao2022,Stromberg2023,Yin2023,Cao2023,Antesberger2024,Tang2024,Richter2026,Guo2026,Qu2026} and on nuclear magnetic resonance platforms~\cite{Nie2022,Xi2024}. Several of its predicted advantages have also been experimentally demonstrated as reviewed in Refs.~\cite{Goswami2020,Rozema2024}. A closely related higher-order operation, the quantum time-flip, coherently controls the input-output direction rather than the order of two operations~\cite{Chiribella2022}. Similarly to the switch, it offers advantages in channel discrimination, quantum communication~\cite{Liu2023b}, quantum metrology~\cite{Agrawal2025}, and has also been realized in photonic experiments~\cite{Guo2024,Stromberg2024}.

As a central resource of quantum information, entanglement underlies numerous tasks ranging from quantum teleportation and secure communication to distributed quantum information processing~\cite{Horodecki2009}. Consequently, understanding how entanglement can be generated, transformed and distributed using coherently controlled quantum processes is a natural question. In the studies regarding the quantum switch and related controlled processes, entanglement has appeared in several conceptually distinct roles, distinguished by when it is introduced and which systems actually share it. Preexisting entanglement between the control system and a remote auxiliary system can enable device-independent certification of indefinite causal order~\cite{vanderLugt2023}. In other works, an entangled state initially shared by more than one target system is transmitted~\cite{Caleffi2020}, distilled~\cite{Zuo2023}, or distributed~\cite{Dey2025} more effectively by coherently controlling the order of relevant operations. Coherent control has also been exploited to generate entanglement among separate target systems or distant network nodes~\cite{Koudia2023,LiuWei2025,Pellitteri2026,Chen2026,Enriquez2026}, and to break the absolute separability of a composite target~\cite{Yanamandra2025}. Lastly, entangling the control registers of two switches realizes entanglement between their temporal orders~\cite{Zych2019,Rubino2022}.

In most studies, the control qubit is treated as an ancillary system, that is, it is typically measured in a suitable basis and the target state is then analyzed conditionally on the outcome, or it is discarded if only the effective target dynamics is of interest. Consequently, the joint state of the control and target before the control is measured or discarded has received comparatively less attention as an object of analysis in its own right. There are, however, a few important exceptions in the recent literature. The operational importance of entanglement across this bipartition was demonstrated by showing that, for the coherently controlled networks of information erasing channels, certain perfect communication and entanglement distribution tasks are possible if and only if the target and control are initially maximally entangled~\cite{Guha2023}. That work treats this entanglement as an initial resource rather than as one generated by the controlled process. In Ref.~\cite{Karpat2026}, although the control was ultimately measured, the preceding joint state of the control and target entered a memory-assisted entropic uncertainty analysis, with the target serving as a quantum memory. In Ref.~\cite{Fellous-Asiani2023}, an energetically constrained quantum switch was considered, and discrimination performance was related to entanglement between the control and remaining systems, including the target. Finally, the control-target entanglement for a class of parameterized switch and time-flip circuits was numerically evaluated in Ref.~\cite{Azado2025}, where comparison with definite causal architectures provided a benchmark for assessing their entanglement generating capabilities. These studies establish the operational relevance of correlations involving the control, either as an initial resource or in the joint output, but they do not provide a systematic and analytical characterization of the entanglement generated for different operations, inputs, and noise parameters. Such a characterization is particularly natural when the available toolbox consists of local operations on the control and target, together with access to a quantum switch or quantum time-flip, but contains no direct entangling gate between them. Starting from a product input, any entanglement generated between the control and target qubits must then arise from the coherent control of alternative target evolutions and therefore directly reflects the entangling capability of the higher order process. The central objective of this manuscript is to determine how much entanglement a coherently controlled quantum process can itself generate between the control and target from an initially uncorrelated input.

In this work, we analytically quantify the entanglement generated between a qubit control and a qubit target by the quantum switch and the quantum time-flip for several canonical families of qubit channels. Rather than measuring or discarding the control, we retain it as one of the two main subsystems and begin with pure product inputs, so that all output entanglement is generated by the coherently controlled process itself. For the quantum switch of two arbitrary qubit unitary channels, we derive exact expressions for the concurrence, its maximum over the control and target inputs, and its Haar average over target inputs for a fixed control state. We give a geometric characterization of the maximal concurrence in terms of the rotation angles of the two unitaries and the relative orientation of their rotation axes. We also establish an exact conservation relation showing how the squared initial control coherence is divided between the squared residual control coherence and the square of the generated concurrence. We show that replacing one unitary by a binary random unitary channel simply rescales the concurrence, its maximum, and its Haar average by the noise parameter, while replacing both produces a rescaling by the product of the two noise parameters.

We then identify a general mechanism which prevents entanglement generation between the control and the target qubits. That is, whenever the two channels commute under composition, their quantum switch cannot entangle any initially separable state of the control and target. This result explains the absence of entanglement when a channel is switched with itself and applies, in particular, to pairs of Pauli channels, pairs of amplitude damping channels, and pairs of generalized amplitude damping channels that share a common stationary state. Dissipation itself, however, does not prevent entanglement generation. Remarkably, switching a completely damping amplitude damping channel with a bit-flip operation can transform a product input into a maximally entangled bipartite state. For two generalized amplitude damping channels, we derive an upper bound determined by the two damping strengths and the difference between their stationary populations. Moreover, we show that a complete relaxation family saturates this bound, and also relaxation toward opposite pure stationary states gives the largest possible concurrence for this pair of channels. Finally, for the quantum time-flip of unitary and binary random unitary channels, we derive exact expressions for the concurrence and its maximum, identify the unitary parameters that permit maximal entanglement, and show that binary mixing reduces the unitary concurrence by a simple factor set by the noise parameter. Together, these results reveal both the mechanisms that enable entanglement between the control and target and the structural conditions that prevent its generation through coherent control of channel order or input and output direction.

This paper is organized as follows. In Sec.~\ref{sec2}, we introduce the quantum switch and the quantum time-flip. In Sec.~\ref{sec3}, we analyze control-target entanglement generated by the quantum switch for unitary and noisy qubit processes. In Sec.~\ref{sec4}, we present the corresponding treatment for the quantum time-flip. Finally, in Sec.~\ref{sec5}, we summarize our results and discuss their implications.

\section{Quantum Switch and Time-Flip}
\label{sec2}

In this section, we introduce the quantum switch and the quantum time-flip as higher-order maps, also known as quantum supermaps. Unlike quantum channels, which transform quantum states, these maps take as inputs the quantum channels themselves. Specifically, the quantum switch is a two-slot supermap whose inputs are two quantum channels. It coherently controls the two orders and constitutes a causally nonseparable higher-order process. The quantum time-flip, on the other hand, takes a bidirectional quantum channel as input and coherently controls its input-output direction.

\subsection{Quantum Switch}
\label{sec2.1}

A quantum channel, representing a physical quantum process, is a completely positive trace-preserving linear map $\mathcal{E}: \mathcal{L}(\mathcal{H}^{\mathrm{in}}) \longrightarrow \mathcal{L}(\mathcal{H}^{\mathrm{out}})$, where $\mathcal{H}^{\mathrm{in}}$ and $\mathcal{H}^{\mathrm{out}}$ denote the input and output Hilbert spaces, respectively, and $\mathcal{L}(\mathcal{H})$ denotes the space of linear operators acting on $\mathcal{H}$. Every such channel admits a Kraus decomposition
\begin{equation}
    \mathcal{E}(\rho) = \sum_i E_i \rho E_i^\dagger,
    \qquad
    \sum_i E_i^\dagger E_i = \mathbb{I}_{\mathrm{in}},
\end{equation}
where $E_i:\mathcal{H}^{\mathrm{in}}\rightarrow\mathcal{H}^{\mathrm{out}}$ are the Kraus operators, $\rho$ is the density operator describing the state of the system, and $\mathbb{I}_{\mathrm{in}}$ is the identity operator acting on $\mathcal{H}^{\mathrm{in}}$.

The quantum switch takes as inputs two channels, $\mathcal{E}$ and $\mathcal{F}$, having Kraus operators $\{E_i\}$ and $\{F_j\}$, respectively. Both channels act on the target system $T$, while a two-level control system $C$ determines their order. The switched process is described by the Kraus operators
\begin{equation}
    S_{ij} = \ket{0}\bra{0}\otimes F_jE_i + \ket{1}\bra{1}\otimes E_iF_j.
    \label{eq:switch-kraus}
\end{equation}
For a bipartite input state $\rho_{CT}^{\mathrm{in}}$ of the control and target, the joint output state is given by
\begin{equation}
    \rho_{CT}
    =
    \sum_{i,j} S_{ij}\rho_{CT}^{\mathrm{in}}S_{ij}^{\dagger}.
\end{equation}
If the control is prepared in $\ket{0}$, the target undergoes the definite order $\mathcal{F}\circ\mathcal{E}$, whereas $\ket{1}$ selects the reverse order $\mathcal{E}\circ\mathcal{F}$. When the control is prepared in a coherent superposition of these basis states, then the two orders are implemented coherently, defining a process having indefinite causal order. Considering the completeness relations for the Kraus operators of $\mathcal{E}$ and $\mathcal{F}$, it is not difficult to see that the switched process is itself completely positive and trace preserving. Although Eq.~\eqref{eq:switch-kraus} is expressed in terms of Kraus operators, the process is independent of the particular Kraus representations chosen for the two channels~\cite{Chiribella2013}. Moreover, among the linear supermaps that preserve complete positivity under arbitrary ancillary extensions, the action of the two-slot quantum switch on arbitrary quantum operations is uniquely determined by its action on unitary operations~\cite{Dong2023}.

\subsection{Quantum Time-Flip}
\label{sec2.2}

Unlike the quantum switch, which takes two channels as inputs and coherently controls their order, the quantum time-flip acts on a single bidirectional channel and coherently controls its input-output direction. A bidirectional channel $\mathcal{E}$ is represented by a bistochastic map, meaning that it is both trace preserving and unital. Its Kraus operators satisfy
\begin{equation}
    \sum_i E_i^\dagger E_i = \sum_i E_i E_i^\dagger = \mathbb{I}.
    \label{eq:bistochastic}
\end{equation}
The first equality guarantees trace preservation, while the second guarantees that the channel preserves the identity operator. With respect to a fixed choice of bases, the reversed input-output direction is represented by the transposed Kraus operators $E_i^{T}$~\cite{Chiribella2022}. The quantum time-flip is described by the Kraus operators
\begin{equation}
    T_i
    =
    \ket{0}\bra{0}\otimes E_i
    +
    \ket{1}\bra{1}\otimes E_i^{T} .
\end{equation}
For a joint control-target input state $\varrho_{CT}^{\mathrm{in}}$, its output is
\begin{equation}
    \varrho_{CT}
    =
    \sum_i T_i\varrho_{CT}^{\mathrm{in}}T_i^\dagger.
\end{equation}
The unitality condition in Eq.~\eqref{eq:bistochastic} ensures that the transposed operators also satisfy the trace-preserving condition. Consequently, when the control is in $\ket{0}$, the channel acts in its original input-output direction, whereas when the control is in $\ket{1}$, it acts in the reversed direction. A coherent superposition of the control states places the two directions in a coherent superposition. As in the case of the quantum switch, the resulting process is independent of the particular Kraus representation chosen for $\mathcal{E}$. Moreover, the quantum time-flip cannot be decomposed as a classical mixture of processes having definite forward or backward input-output direction, meaning that it has indefinite input-output direction~\cite{Chiribella2022}.

\section{Entangling capability of the quantum switch}
\label{sec3}

\subsection{Unitary Channels}
\label{sec:unitary-switch}

Let us first consider the quantum switch of two unitary qubit channels,
$\mathcal{U}(\rho)=U\rho U^\dagger$ and $\mathcal{V}(\rho)=V\rho V^\dagger$. The corresponding switch operator is given by
\begin{equation}
    S_{UV}
    =
    \ket{0}\bra{0}\otimes VU
    +
    \ket{1}\bra{1}\otimes UV,
\end{equation}
where the computational basis of the control labels the two alternative orders in which $U$ and $V$ act on the target, corresponding to $\mathcal{E}=\mathcal{U}$ and $\mathcal{F}=\mathcal{V}$ in Eq.~\eqref{eq:switch-kraus}. We take the initial control state as 
\begin{equation}
    \ket{\psi_C}
    =
    \alpha\ket{0}
    +
    \beta\ket{1},
    \qquad
    |\alpha|^2+|\beta|^2=1,
    \label{eq:control-input}
\end{equation}
and the target is prepared in an arbitrary pure state $\ket{\psi}$. Acting with $S_{UV}$ on the product input $\ket{\psi_C}\otimes\ket{\psi}$ yields
\begin{align}
    \ket{\Psi}
    &=
    \alpha\ket{0}\otimes VU\ket{\psi}
    +
    \beta\ket{1}\otimes UV\ket{\psi}
    \nonumber\\
    &=
    \alpha\ket{0}\otimes\ket{\phi_0}
    +
    \beta\ket{1}\otimes\ket{\phi_1},
    \label{eq:unitary-switch-output}
\end{align}
where we introduced $\ket{\phi_0}=VU\ket{\psi}$ and $\ket{\phi_1}=UV\ket{\psi}$, denoting the two alternatively ordered evolutions of the target. The overlap of these two states,
\begin{equation}
    s
    =
    \braket{\phi_1|\phi_0}
    =
    \bra{\psi}X\ket{\psi},
    \qquad
    X = V^\dagger U^\dagger VU,
\end{equation}
quantifies how distinguishable the two ordered evolutions are. Since $X$ is a product of unitaries it is itself unitary, so $|s|\le 1$ for every target state $\ket{\psi}$. The controlled-gate structure of the induced unitary can be made explicit as
\begin{equation}
    S_{UV}
    =
    \left(\mathbb{I}\otimes UV\right)
    \left(
        \ket{0}\!\bra{0}\otimes X
        +
        \ket{1}\!\bra{1}\otimes\mathbb{I}
    \right).
\end{equation}
The factor $\mathbb{I} \otimes UV$ acts locally only on the target and does not affect the control-target entanglement. This actually means that $S_{UV}$ is locally equivalent to a controlled-unitary gate with target operation $X$, belonging to a well-studied class of gates~\cite{zanardi2000,kraus2001,wang2003}. In the present setting, its maximal entangling capability is fixed by the spectrum of $X$ alone, as Eqs.~\eqref{eq:concurrence-s}
and~\eqref{eq:max-concurrence-spectral} make explicit.
Regardless, this correspondence does not indicate the equivalence of physical resources, since in our case $U$ and $V$ act only on the target and no direct control-target entangling interaction is independently supplied. The switch specific content is that $X = V^{\dagger} U^{\dagger} V U$ is fixed by the two alternative orders of the inserted unitary channels, leading to the characterization presented below. In fact, this mathematical correspondence is specific to the unitary channel case and does not generally extend to noisy channels.

The state in Eq.~\eqref{eq:unitary-switch-output} is a pure state, and it is separable if and only if either $\alpha\beta=0$, so that the control is in a computational basis state, or $\ket{\phi_0}$ and $\ket{\phi_1}$ coincide up to a phase, i.e., $\ket{\phi_0}=e^{i\chi}\ket{\phi_1}$ for some $\chi\in\mathbb{R}$, equivalently $|s|=1$. Entanglement is then generated when
\begin{equation}
    \alpha\beta\neq 0
    \quad\text{and}\quad
    |s|<1 .
    \label{eq:ent-condition}
\end{equation}
It is straightforward to identify the unitary channel pairs $(U,V)$ for which the switch generates no entanglement on any target, i.e., $|s|=1$ for all  $\ket{\psi}$. This simply requires that $X=c\mathbb{I}$ for some constant $c$. As the eigenvalues of a unitary matrix are phases, so is its determinant, and thus $\det X=(\det V)^\ast(\det U)^\ast(\det V)(\det U)=1$ for any $U$ and $V$, so that $X\in SU(2)$. Then, $c^2=1$ and $c=\pm1$. If $X=\mathbb{I}$, we have $[U,V]=0$ and $s=1$, and the output is $(\alpha\ket{0}+\beta\ket{1})\otimes\ket{\phi_0}$. If $X=-\mathbb{I}$, we have $\{U,V\}=0$ and $s=-1$, and the output is $(-\alpha\ket{0}+\beta\ket{1})\otimes\ket{\phi_1}$. We see that if $U$ and $V$ commute, the two ordered target states are identical, $\ket{\phi_0}=\ket{\phi_1}$, while when they anticommute, they differ solely by a simple global phase, $\ket{\phi_0}=-\ket{\phi_1}$. The output is separable in both cases, showing that non-commutativity of the unitary operators is necessary but not sufficient for entanglement generation.

To quantify the control-target entanglement generated from an arbitrary pure product input, we utilize the concurrence~\cite{Wootters1998}. Since the switch of two unitary channels maps a pure input to a pure two-qubit output, the concurrence is given by $\mathcal{C}=2\sqrt{\det\rho_C}$,
where $\rho_C=\mathrm{Tr}_T\rho_{CT}$ is the reduced density operator of the control. Tracing out the target qubit from $\rho_{CT}=\ket{\Psi}\bra{\Psi}$ gives
\begin{equation}
    \rho_C
    =
    \begin{pmatrix}
        |\alpha|^2 & \alpha\beta^\ast s\\
        \alpha^\ast\beta\, s^\ast & |\beta|^2
    \end{pmatrix},
\end{equation}
whose determinant is $\det\rho_C=|\alpha|^2|\beta|^2(1-|s|^2)$, so that
\begin{equation}
    \mathcal{C}(\rho_{CT})
    =
    2|\alpha\beta|\sqrt{1-|s|^2},
    \label{eq:concurrence-s}
\end{equation}
which is in accord with the condition given in Eq.~\eqref{eq:ent-condition}. It is clear that for fixed $U$, $V$, and $\ket{\psi}$, the concurrence is maximized by a balanced control state, $|\alpha|=|\beta|=1/\sqrt{2}$.

In fact, the same overlap $s$, controlling the concurrence generated between the control and the target qubits, also determines the residual coherence of the control, reflecting the broader operational connection between these two resources~\cite{Streltsov2015}. For the $\ell_1$-norm of coherence in the computational basis, which equals twice the modulus of the off-diagonal element of a qubit density operator~\cite{Baumgratz2014}, the initial and final control coherences are given by
\begin{equation}
    C_{\ell_1}(\rho_C^{\mathrm{in}})=2|\alpha\beta|,
    \qquad
    C_{\ell_1}(\rho_C)=2|\alpha\beta||s|,
\end{equation}
respectively. Combining these expressions with Eq.~\eqref{eq:concurrence-s}, we obtain a conservation relation between entanglement and coherence for the switch of two unitary channels,
\begin{equation}
    C_{\ell_1}(\rho_C^{\mathrm{in}})^2
    =
    C_{\ell_1}(\rho_C)^2 + \mathcal{C}(\rho_{CT})^2.
    \label{eq:coherence-concurrence-l1}
\end{equation}
The initial control coherence is shared between the residual control coherence and the control-target concurrence. For a fixed initial control state, the concurrence reaches its maximum when the final control state is incoherent. We also note that connections between coherence and entanglement have been previously discussed in the context of complementarity for pure bipartite systems~\cite{Jakob2010}. 

The dependence of concurrence on the target state can be made explicit by expressing $\ket{\psi}$ in the eigenbasis of $X$. As $X\in SU(2)$, its eigenvalues form a conjugate pair,
\begin{equation}
    X\ket{v_\pm}
    =
    e^{\pm i\theta_X/2}\ket{v_\pm},
    \qquad
    0\leq\theta_X\leq 2\pi,
\end{equation}
where $\ket{v_+}$ and $\ket{v_-}$ are the two orthonormal eigenstates. Expanding the target state in this basis as
\begin{equation}
    \ket{\psi}
    =
    \cos(\vartheta/2)\,\ket{v_+}
    +
    e^{i\varphi}\sin(\vartheta/2)\,|v_-\rangle,
\end{equation}
where $0\leq\vartheta\leq\pi$ and $0\leq\varphi<2\pi$, the overlap is
\begin{align}
    s
    &=
    \cos^2(\vartheta/2)e^{i\theta_X/2}
    +
    \sin^2(\vartheta/2)e^{-i\theta_X/2} \nonumber\\
    &=
    \cos(\theta_X/2)
    +
    i\cos(\vartheta)\sin(\theta_X/2).
\end{align}
Then, the concurrence takes the form
\begin{equation}
   \mathcal{C} = 2|\alpha\beta| \left|\sin(\theta_X/2)\right|\sin(\vartheta).
    \label{eq:concurrence-spectral}
\end{equation}
For $X\neq\pm\mathbb{I}$ and $\alpha\beta\neq0$, the concurrence vanishes
if and only if the target state is an eigenstate of $X$, corresponding to
$\vartheta=0$ or $\vartheta=\pi$. Considering fixed $U$ and $V$, the concurrence is maximized over the target state when $\vartheta=\pi/2$, corresponding to the target state family
\begin{equation}
    \ket{\psi_{\max}}
    =
    \frac{\ket{v_+}+e^{i\varphi}\ket{v_-}}{\sqrt{2}}.
\end{equation}
Maximizing also over the control, with the maximum attained when $|\alpha|=|\beta|=1/\sqrt{2}$, gives
\begin{equation}
    \mathcal{C}_{\max}(U,V)
    =
    \max_{\ket{\psi_C},\ket{\psi}} \mathcal{C}
    =
    |\!\sin(\theta_X/2)|.
    \label{eq:max-concurrence-spectral}
\end{equation}
Hence, the switch is able to generate a maximally entangled state if and only if $\theta_X=\pi$, equivalently $\mathrm{Tr}\,X=0$, i.e., the eigenvalues of $X$ are $\{i,-i\}$, which is the unique antipodal eigenvalue pair compatible with $\det X=1$.

To characterize the concurrence generated without selecting a particular input target state, we next average the concurrence over Haar-distributed pure target states, while keeping $U$, $V$, and the control state fixed. Following the angular parametrization, the corresponding uniform measure on the Bloch sphere is $d\mu(\psi)=\sin(\vartheta)\,d\vartheta\,d\varphi /4\pi$. Using Eq.~\eqref{eq:concurrence-spectral}, the averaged concurrence is given by
\begin{align}
   \overline{\mathcal C}_{\psi}
    &=
    \frac{2|\alpha\beta|}{4\pi}\left|\sin(\theta_X/2)\right|\int_0^{2\pi}d\varphi\int_0^\pi\sin^2(\vartheta)d\vartheta
    \nonumber\\
    &=
    \frac{\pi}{2}|\alpha\beta| \left|\sin(\theta_X/2)\right|.
\end{align}
Recalling Eq.~\eqref{eq:max-concurrence-spectral}, the averaged concurrence becomes
\begin{equation}
    \overline{\mathcal C}_{\psi}
    =
    \frac{\pi}{2}|\alpha\beta| \mathcal C_{\max}(U,V),
\end{equation}
and for a maximally coherent control, $|\alpha|=|\beta|=1/\sqrt{2}$, it simply reduces to the relation
\begin{equation}
    \overline{\mathcal C}_{\psi}
    =
    \frac{\pi}{4}\mathcal C_{\max}(U,V).
\end{equation}
Interestingly, for the switch of two qubit unitary channels with a maximally coherent control, the proportionality factor $\pi/4$ is universal, being independent of the particular pair $U$ and $V$. Thus, the Haar-averaged and maximal concurrences induce the same ordering of unitary pairs, that is, optimizing $\mathcal{C}_{\max}(U,V)$ simultaneously optimizes the average entangling capability over pure target states. We also note that a recent concurrence-based study of controlled-phase gates considered a different setting, where both product inputs are Haar-averaged~\cite{rudzinski2026}.

\begin{figure}[t]
\centering
\includegraphics[width=0.92\columnwidth]{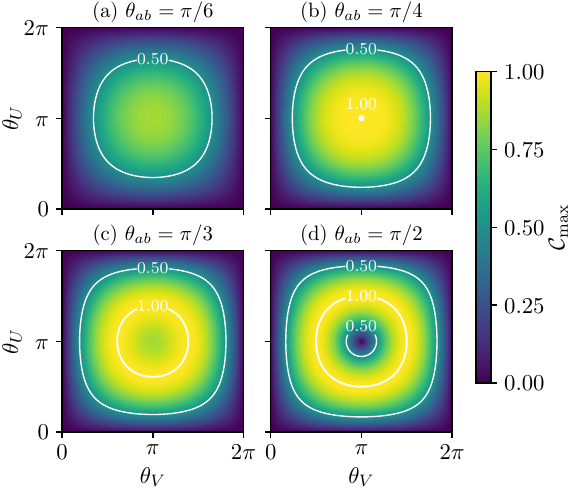}
\caption{Density plots of the maximal concurrence for switch of two unitary channels, $\mathcal{C}_{\max}(U,V)$, as a function of the rotation angles $\theta_U$ and $\theta_V$ for four different relative angles between the rotation axes. Here, the white point and curves highlight the contours on which the concurrence has constant value.}
\label{fig1}
\end{figure}

The maximal concurrence admits a closed form expression also in terms of the geometric parameters of the two unitaries, namely, their rotation angles and the relative orientation of their rotation axes. As the global phases of $U$ and $V$ cancel in $X$, one can replace them by $SU(2)$ matrices,
$\widetilde U=(\det U)^{-1/2}U$ and $\widetilde V=(\det V)^{-1/2}V$, without changing $\mathcal{C}_{\max}$. Dropping the tildes, we can express them in axis-angle
form as
\begin{align}
    U&=\cos(\theta_U/2)\,\mathbb{I}-i\sin(\theta_U/2)\,(\hat{a}\cdot\vec{\sigma}),
    \nonumber\\
    V&=\cos(\theta_V/2)\,\mathbb{I}-i\sin(\theta_V/2)\,(\hat{b}\cdot\vec{\sigma}),
    \label{eq:UV-axisangle}
\end{align}
where $\theta_U,\theta_V\in[0,2\pi]$ are the rotation angles, $\hat{a}$ and $\hat{b}$ are unit vectors along the rotation axes, and $\vec{\sigma}=(\sigma_x,\sigma_y,\sigma_z)$ is the vector of Pauli operators. Denoting the angle between the two axes by $\theta_{ab}\in[0,\pi]$, so that
$\hat{a}\cdot\hat{b}=\cos\theta_{ab}$, a direct evaluation of $X$ yields (see
Appendix~\ref{appA})
\begin{equation}
    \frac{\mathrm{Tr}\,X}{2}
    =
    1-2\sin^2(\theta_U/2)\sin^2(\theta_V/2)\sin^2(\theta_{ab}).
    \label{eq:trace-axisangle}
\end{equation}
Since $\mathrm{Tr}\,X=2\cos(\theta_X/2)$, Eqs.~\eqref{eq:max-concurrence-spectral} and~\eqref{eq:trace-axisangle} give
\begin{equation}
    \mathcal{C}_{\max}(U,V)
    =
    2\sqrt{t(1-t)},
    \label{eq:cmax-axisangle}
\end{equation}
with $t=\sin^2(\theta_U/2)\sin^2(\theta_V/2)\sin^2(\theta_{ab})$, and $ 0\leq t\leq1$. This identifies all entangling and nonentangling combinations of the unitary pair $U$ and $V$. The maximal concurrence vanishes if $\theta_U\in\{0,2\pi\}$ or $\theta_V\in\{0,2\pi\}$, that is, when either operation is proportional to the identity, and if $\theta_{ab}=0$ or $\pi$, i.e., when the two rotation axes are collinear and thus the operations commute. It also vanishes when $\theta_U=\theta_V=\pi$ with $\theta_{ab}=\pi/2$, corresponding to $t=1$, which is the anticommuting case. On the other hand, the switch is maximally entangling, $\mathcal{C}_{\max}=1$, if and only if $t=1/2$. As $t\leq\sin^2(\theta_{ab})$, a maximally entangled output is attainable only for $\pi/4\leq\theta_{ab}\leq3\pi/4$. The resulting dependence of $\mathcal{C}_{\max}$ on the two rotation angles is shown in Fig.~\ref{fig1}. The maximal entanglement set evolves from a single point at $\theta_{ab}=\pi/4$ into closed curves for the larger relative angles. For $\theta_{ab}=\pi/2$, this curve surrounds the nonentangling point $(\theta_U,\theta_V)=(\pi,\pi)$.

\subsection{Binary Random Unitary Channels}
\label{sec:binary-unitary-switch}

We now consider the simplest unital noisy extension of the preceding setting by replacing one of the unitary channels with the binary random unitary channel
\begin{equation}
    \mathcal{R}_{p}(\rho)
    =
    (1-p)\rho+pV\rho V^\dagger,
    \qquad
    0\leq p\leq1,
\end{equation}
where $p$ is the noise parameter, representing the probability with which the unitary error $V$ is applied. A Kraus representation of this channel is given by
\begin{equation}
    K_0=\sqrt{1-p}\,\mathbb{I},
    \qquad
    K_1=\sqrt{p}\,V.
\end{equation}
If we switch $\mathcal{R}_{p}$ with the unitary channel $\mathcal{U}(\rho)=U\rho U^\dagger$, the corresponding Kraus operators take the form
\begin{equation}
    S_i
    =
    \ket{0}\bra{0}\otimes K_iU
    +
    \ket{1}\bra{1}\otimes UK_i,
    \quad
    i=0,1.
\end{equation}
For the same product input $\ket{\psi_C}\otimes\ket{\psi}$ considered above, the output state after the switch can be written as
\begin{equation}
    \rho_{CT}
    =
    \ket{w_0}\bra{w_0}
    +
    \ket{w_1}\bra{w_1},
    \label{eq:binary-ru-output}
\end{equation}
where the unnormalized vectors read
\begin{align}
    \ket{w_0}
    &=
    \sqrt{1-p} \ket{\psi_C}\otimes U\ket{\psi}
    \equiv
    \sqrt{1-p}\ket{\chi}, \nonumber\\
    \ket{w_1}
    &=
    \sqrt{p} \ket{\Psi}.
    \label{eq:binary-ru-vectors}
\end{align}
Here, $\ket{\Psi}$ is the unitary switch output given in Eq.~\eqref{eq:unitary-switch-output}, and
$\ket{\chi}=\ket{\psi_C}\otimes U\ket{\psi}$ is a product state. Then,
\begin{equation}
    \rho_{CT}
    =
    (1-p)\ket{\chi}\bra{\chi}
    +
    p\ket{\Psi}\bra{\Psi}.
    \label{eq:binary-ru-mixture}
\end{equation}
The concurrence can be evaluated from the singular values of the spin-flip overlap matrix. For the state given in Eq.~\eqref{eq:binary-ru-mixture}, whose rank is at most two, we have
\begin{align}
    \mathcal{C}(\rho_{CT})
    =
    p\,\mathcal{C}\left(\ket{\Psi}\bra{\Psi}\right)
    =
    2p|\alpha\beta|\sqrt{1-|s|^2},
    \label{eq:binary-ru-concurrence}
\end{align}
as shown in Appendix~\ref{appB}. As a consequence, entanglement is generated between the control and target if and only if
\begin{equation}
    p>0,
    \qquad
    \alpha\beta\neq0,
    \qquad
    |s|<1.
\end{equation}
Thus, for every $p>0$, the conditions on the control state, target state, and unitary pair are the same as in the two unitary channel case. The net effect of the binary random unitary channel is to rescale the concurrence by $p$. The control and target states maximizing the concurrence are also unchanged, and the maximal concurrence becomes
\begin{equation}
    \mathcal{C}_{\max}(U,V,p)
    =
    p\,\mathcal{C}_{\max}(U,V)
    =
    2p\sqrt{t(1-t)}.
\end{equation}
The same linear scaling holds under Haar averaging. In particular, for a maximally coherent control,
\begin{equation}
    \overline{\mathcal{C}}_{\psi}(U,V,p)
    =
    p\,\overline{\mathcal{C}}_{\psi}(U,V,1)
    =
    \frac{\pi}{4}
    \mathcal{C}_{\max}(U,V,p).
\end{equation}
Then, the universal proportionality between the averaged and maximal concurrences remains unchanged. More importantly, Eq.~\eqref{eq:binary-ru-concurrence} shows that the switch can generate control-target entanglement even in the genuinely noisy regime $0<p<1$. In particular, for every $0<p<1$, the concurrence is nonzero whenever $\alpha\beta\neq0$ and $|s|<1$. Considering fixed $U$, $V$, and fixed input states satisfying these conditions, the generated concurrence increases linearly with the noise parameter $p$. At $p=0$, $\mathcal{R}_{p}$ reduces to the identity channel and no entanglement is generated. On the other hand, if $p=1$, $\mathcal R_p$ reduces to the unitary channel $\mathcal V$, and the unitary switch result is recovered.

\begin{figure}[t]
\centering
\includegraphics[width=\columnwidth]{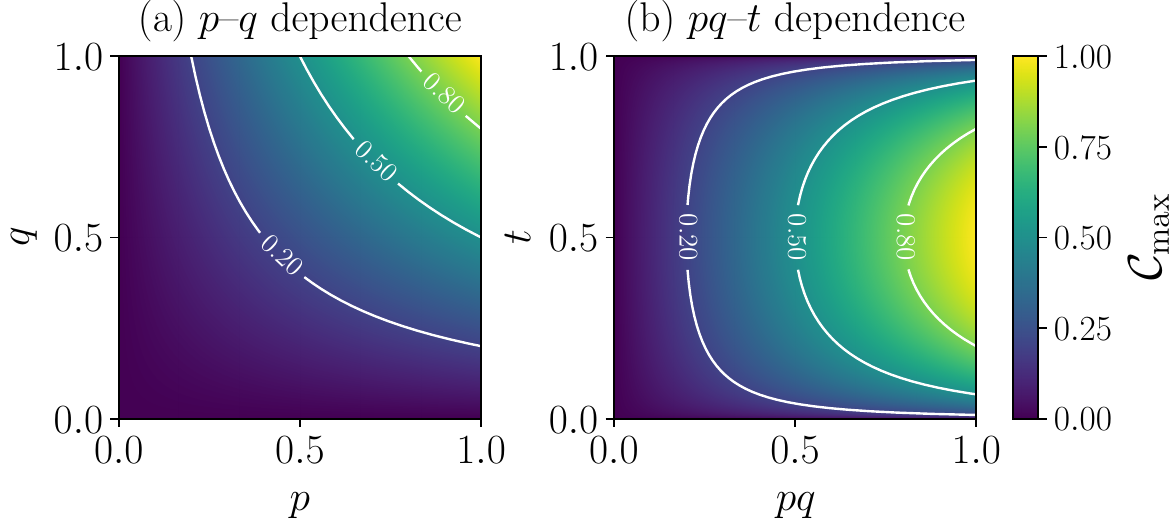}
\caption{Density plots of the maximal concurrence generated via quantum switch of two binary random unitary channels. (a) Dependence of concurrence on the noise parameters $p$ and $q$ for $t=1/2$ and (b) on the parameter $pq$ and the unitary parameter $t$. The white contours indicate the parameter values corresponding to $\mathcal{C}_{\max}=0.20$, $0.50$, and $0.80$.}
\label{fig:two-binary-ru}
\end{figure}

The previous case considered a binary random unitary channel switched with a unitary channel, for which the concurrence is obtained by a simple rescaling of the unitary-switch result. We now extend this analysis to the case where both channels entering the switch are binary random unitary channels. We consider
\begin{align}
    \mathcal{R}_{p,U}(\rho)
    &=
    (1-p)\rho+pU\rho U^\dagger, \nonumber\\
    \mathcal{R}_{q,V}(\rho)
    &=
    (1-q)\rho+qV\rho V^\dagger ,
\end{align}
where $p,q \in [0,1]$ denote the noise parameters for the considered channels. The Kraus operators are
\begin{align}
    K_0=\sqrt{1-p}\,\mathbb{I},
    \qquad
    K_1=\sqrt{p}\,U,
    \nonumber\\
    L_0=\sqrt{1-q}\,\mathbb{I},
    \qquad
    L_1=\sqrt{q}\,V .
\end{align}
For the same initial product input state considered previously, the switched output state can be expressed as
\begin{equation}
    \rho_{CT}
    =
    \sum_{i,j=0}^{1}
    \ket{w_{ij}}\bra{w_{ij}},
    \label{eq:two-binary-ru-output}
\end{equation}
where 
\begin{equation}
    \ket{w_{ij}}
    =
    \left( \alpha\ket{0}\otimes L_jK_i\ket{\psi}
    +
    \beta\ket{1}\otimes K_iL_j\ket{\psi} \right).
\end{equation}
The four vectors appearing in Eq.~\eqref{eq:two-binary-ru-output} take the form
\begin{align}
    \ket{w_{00}}
    &=
    \sqrt{(1-p)(1-q)}
    \ket{\psi_C}\otimes\ket{\psi},
    \nonumber\\
    \ket{w_{01}}
    &=
    \sqrt{(1-p)q}
    \ket{\psi_C}\otimes V\ket{\psi}
    \nonumber\\
    \ket{w_{10}}
    &=
    \sqrt{p(1-q)}
    \ket{\psi_C}\otimes U\ket{\psi},
    \nonumber\\
    \ket{w_{11}}
    &=
    \sqrt{pq}\ket{\Psi},
   \label{eq:two-binary-ru-vectors}
\end{align}
where $\ket{\Psi}$ is the unitary switch output given in Eq.~\eqref{eq:unitary-switch-output}. While the first three terms correspond to product states of the control and target, the last one contains the two alternatively ordered evolutions of the target generated by the unitary pair $U$ and $V$. Utilizing the spin-flip overlap matrix introduced in Appendix~\ref{appB}, the overlaps involving the first three product states appear identically in the two singular values, so their difference is determined only by the spin-flip self-overlap of $\ket{w_{11}}$. The control-target concurrence then reduces to
\begin{equation}
    \mathcal{C}(\rho_{CT})
    =
    pq\,
    \mathcal{C}\left(\ket{\Psi}\bra{\Psi}\right),
\end{equation}
as explicitly shown in Appendix~\ref{appB}, and we have
\begin{equation}
    \mathcal{C}(\rho_{CT})
    =
    2pq|\alpha\beta| \sqrt{1-|s|^2}.
    \label{eq:two-binary-ru-concurrence-explicit}
\end{equation}
We see that the overlap between the two ordered target evolutions continues to fully determine the dependence on the unitary pair, exactly as in the unitary case. The effect of replacing both unitary channels by binary random unitary channels is an additional multiplicative factor $pq$. The maximal concurrence reads
\begin{equation}
    \mathcal{C}_{\max}(U,V,p,q)
    =
    pq\,\mathcal{C}_{\max}(U,V)
    =
    2pq\sqrt{t(1-t)} .
\end{equation}
The dependence of the maximal concurrence on the two parameters and the two unitaries is illustrated in Fig.~\ref{fig:two-binary-ru}. The optimal control and target states remain unchanged, since $p$ and $q$ only determine the overall reduction of the concurrence. As consistency checks, for $q=1$, Eq.~\eqref{eq:two-binary-ru-concurrence-explicit} reduces to the result for a binary random unitary channel and a unitary channel in Eq.~\eqref{eq:binary-ru-concurrence}, while for $p=q=1$ it recovers the unitary switch result in Eq.~\eqref{eq:concurrence-s}.

Finally, an important particular case is the self-switch, in which two uses of the same channel are placed in the quantum switch. For an initially factorized control-target input, the control-target output is known to be separable for any quantum channel~\cite{Chiribella2021a}. The binary random unitary family discussed here illustrates this result directly. In the identical channel limit, $p=q$ and $U=V$, the two ordered target evolutions appearing in $\ket{w_{11}}$ coincide,
\begin{equation}
    VU\ket{\psi}=UV\ket{\psi}=U^2\ket{\psi},
\end{equation}
and hence $s=1$. Equation~\eqref{eq:two-binary-ru-concurrence-explicit} then gives
\begin{equation}
    \mathcal C(\rho_{CT})
    =
    2p^2|\alpha\beta|\sqrt{1-|s|^2}
    =0.
\end{equation}
As a consequence, even though two different binary random unitary channels can generate control-target entanglement, their identical channel self-switch cannot, which is in agreement with the general result.

\subsection{Pauli Channels}
\label{sec:pauli-switch}

We next turn to the Pauli channels, which form an important structured subclass of random unitary channels. They extend the binary random unitary model considered above by allowing a mixture of the identity and the three Pauli operations, while their closed operator algebra permits further analytical simplifications. Actually, Pauli channels involve the standard bit-flip, phase-flip, bit-phase-flip, and depolarizing channels as special cases. First, we consider the switch of the
Pauli channel
\begin{equation}
    \mathcal{P}(\rho)
    =
    p_0\rho
    +
    p_x\sigma_x\rho\sigma_x
    +
    p_y\sigma_y\rho\sigma_y
    +
    p_z\sigma_z\rho\sigma_z,
\end{equation}
where $p_0+p_x+p_y+p_z=1$, and $p_\mu\geq0$, with the unitary Hadamard channel described by
\begin{equation}
    \mathcal{H}(\rho)
    =
    H\rho H,
    \qquad
    H
    =
    (\sigma_x+\sigma_z)/\sqrt{2}.
\end{equation}
A Kraus representation of the Pauli channel is given by
\begin{equation}
    K_\mu
    =
    \sqrt{p_\mu}\,\sigma_\mu,
    \quad
    \sigma_0=\mathbb{I},
    \quad
    \mu=0,x,y,z.
\end{equation}
Taking $\mathcal{E}=\mathcal{H}$ and $\mathcal{F}=\mathcal{P}$ in
Eq.~\eqref{eq:switch-kraus}, the Kraus operators of the switched process read
\begin{equation}
    S_\mu
    =
    \ket{0}\bra{0}\otimes K_\mu H
    +
    \ket{1}\bra{1}\otimes HK_\mu,
    \quad
    \mu=0,x,y,z.
\end{equation}
For the same pure separable input state $\ket{\psi_C}\ket{\psi}$, the switched output can be written as
\begin{equation}
    \rho_{CT}
    =
    \sum_{\mu=0,x,y,z}
    \ket{w_\mu}\bra{w_\mu},
\end{equation}
where the four unnormalized vectors are
\begin{equation}
    \ket{w_\mu}
    =
    \sqrt{p_\mu}
    \left(
    \alpha\ket{0}\otimes\sigma_\mu H\ket{\psi}
    +
    \beta\ket{1}\otimes H\sigma_\mu\ket{\psi}
    \right).
    \label{eq:pauli-H-vectors}
\end{equation}
As we prove in Appendix~\ref{appC}, the concurrence for an arbitrary target initial state satisfies
\begin{equation}
    \mathcal{C}(\rho_{CT})
    \leq
    2|\alpha\beta|
    |p_x-p_z|.
    \label{eq:pauli-H-upper-bound}
\end{equation}
This bound is saturated when the target system is prepared in an eigenstate of the Hadamard operator,
\begin{equation}
    H\ket{h_+}
    =
    \ket{h_+}.
\end{equation}
For the above target state, the singular values of the $4\times4$ spin-flip overlap matrix associated with Eq.~\eqref{eq:pauli-H-vectors} are
\begin{align}
    \mu_{1,2}
    &=
    |\alpha\beta|\left[\sqrt{1-(p_0-p_y)^2}\pm|p_x-p_z|
    \right],
\end{align}
and $\mu_3=\mu_4=0$. It simply follows that
\begin{equation}
    \mathcal{C}(\rho_{CT})
    =
    \mu_1-\mu_2
    =
    2|\alpha\beta|
    |p_x-p_z|,
\end{equation}
which clearly saturates Eq.~\eqref{eq:pauli-H-upper-bound}. As a result, for a maximally coherent control state, we have
\begin{equation}
    \mathcal{C}_{\max}\left(\mathcal{H},\mathcal{P}\right)
    =
    |p_x-p_z|.
    \label{eq:pauli-H-maximum}
\end{equation}
It can be easily seen that the maximal concurrence generated by the switch in this case is determined solely by the imbalance between the $\sigma_x$ and $\sigma_z$ components of the Pauli channel. Although $p_0$ and $p_y$ enter the two nonzero singular values separately, their contributions cancel from $\mu_1-\mu_2$. In particular, when $p_x=p_z$, no control-target entanglement is generated for any target initial state for the switched Hadamard and Pauli channels.

\begin{figure}[t]
\centering
\includegraphics[width=\columnwidth]{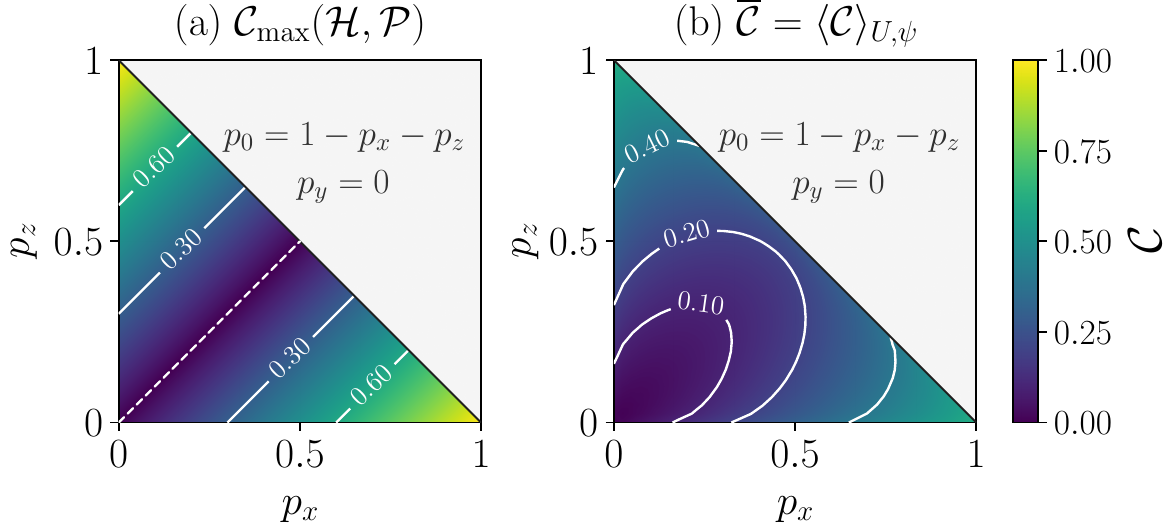}
\caption{Density plots of the concurrence generated by the switch for the two-parameter family of Pauli channels $\mathcal{P}$ and a unitary channel. The physically valid parameter region satisfies $p_x+p_z\leq1$ and $p_x,p_z\geq0$. (a) Maximal concurrence for the Hadamard-Pauli switch. The dashed white line $p_x=p_z$ marks the nonentangling condition. (b) Haar-averaged concurrence for the switch of unitary and Pauli channels, with $U$ and $\ket{\psi}$ sampled independently according to the Haar measure and the control fixed to the maximally coherent state.}
\label{fig3}
\end{figure}

To illustrate this result, in Fig.~\ref{fig3}, we consider the two-parameter family of Pauli channels defined by $p_y=0$. In this case, $p_0=1-p_x-p_z$, and the valid parameter region is the triangle defined by $p_x,p_z\geq0$ and $p_x+p_z\leq1$. Fig.~\ref{fig3}(a) displays the exact optimized result in Eq.~\eqref{eq:pauli-H-maximum}. The maximal concurrence reaches unity at the pure bit-flip and phase-flip vertices, $(p_x,p_z)=(1,0)$ and $(0,1)$, respectively, and vanishes along the dashed line $p_x=p_z$. For comparison, in panel~(b) we replace $\mathcal H$ by a general unitary channel $\mathcal U(\rho)=U\rho U^\dagger$ and consider the Haar-averaged concurrence $\overline{\mathcal C}(p_x,p_z)=\left\langle\mathcal C(\rho_{CT})\right\rangle_{U,\psi}$. Here, the unitary matrix $U$ and the input $\ket{\psi}$ are sampled independently according to the Haar measure, with the control state fixed to the maximally coherent state. As demonstrated by the numerical results in Fig.~\ref{fig3}(b), $\overline{\mathcal C}$ is generally nonzero along $p_x=p_z$, except at the identity vertex $(p_x,p_z)=(0,0)$, so the nonentangling line in Fig.~\ref{fig3}(a) is not a generic feature of the unitary-Pauli switch.

We finalize this subsection by considering the switch of two Pauli channels, taking $\mathcal{E}=\mathcal{P}$ and $\mathcal{F}=\mathcal{Q}$, where
\begin{equation}
    \mathcal{Q}(\rho)
    =
    \sum_{\nu=0,x,y,z}
    q_\nu\sigma_\nu\rho\sigma_\nu,
    \quad
    q_\nu\geq0,
    \quad
    \sum_\nu q_\nu=1.
\end{equation}
For the pure product input $\ket{\psi_C}\ket{\psi}$, each Kraus contribution to the switched output can be written as
\begin{align}
    \ket{w_{\mu\nu}}
    &=
    \sqrt{p_\mu q_\nu}\left(
        \alpha\ket{0}\otimes\sigma_\nu\sigma_\mu\ket{\psi}
        +
        \beta\ket{1}\otimes\sigma_\mu\sigma_\nu\ket{\psi} \right)
    \nonumber\\
    &=
    \sqrt{p_\mu q_\nu}
    \left(\alpha\ket{0}
        +
        \eta_{\mu\nu}\beta\ket{1}\right)
    \otimes
    \sigma_\nu\sigma_\mu\ket{\psi},
\end{align}
where
$\sigma_\mu\sigma_\nu
=\eta_{\mu\nu}\sigma_\nu\sigma_\mu$
with $\eta_{\mu\nu}=\pm1$. Therefore, every Kraus contribution is a product state, and their incoherent sum is separable. Consequently, $\mathcal{C}(\rho_{CT})=0$ for arbitrary choice of noise parameters, control states, and target states considered here, even if the two Pauli channels are different. For the maximally coherent control input $\ket{+}=(\ket{0}+\ket{1})/\sqrt{2}$, commuting Pauli pairs leave the control in $\ket{+}$, whereas anticommuting pairs transform it into $\ket{-}=(\ket{0}-\ket{1})/\sqrt{2}$, which is a structure previously explored in communication through switched Pauli channels~\cite{Bhattacharya2021}. This contrasts with the Hadamard-Pauli switch we explored. For two Pauli channels, reversing the order of any Kraus pair changes the target operation only by an overall sign, so the two control branches always contain the same target state up to a phase. In the Hadamard-Pauli case, however, the $\sigma_x$ branch produces $\sigma_xH\ket{\psi}$ and
$H\sigma_x\ket{\psi}=\sigma_zH\ket{\psi}$ in the two orders, while the $\sigma_z$ branch produces the same two target states in the reversed order. These states are generally not proportional, allowing control-target entanglement to be generated.

\subsection{Channel Commutation and Separability}
\label{sec:channel-commutation}

The separability of the switch of two Pauli channels, together with the self-switch result and the nonentangling unitary cases identified above, points to a common condition that is independent of the particular channel family. We now formulate this condition directly in terms of the two composed channels. Since the quantum switch coherently controls the ordered compositions $\mathcal{F}\circ\mathcal{E}$ and $\mathcal{E}\circ\mathcal{F}$, we show that their equality as maps
forces the switched channel to admit a product Kraus representation. Consequently, the switched channel cannot generate control-target entanglement from a
separable input.

To see this, let us consider two quantum channels acting on the same target
system,
\begin{equation}
    \mathcal{E}(\rho)
    =
    \sum_{i=1}^{m}E_i\rho E_i^\dagger,
    \qquad
    \mathcal{F}(\rho)
    =
    \sum_{j=1}^{n} F_j\rho F_j^\dagger.
\end{equation}
We define the Kraus operators of the two alternatively ordered compositions as
\begin{equation}
    A_{ij}=F_jE_i,
    \qquad
    B_{ij}=E_iF_j.
\end{equation}
Writing $k=(i,j)$ as a single label, both Kraus families $\{A_k\}$ and $\{B_k\}$ contain $N=mn$ operators. The switch Kraus operators in Eq.~\eqref{eq:switch-kraus} can then be written as
\begin{equation}
    S_k
    =
    \ket{0}\bra{0}\otimes A_k
    +
    \ket{1}\bra{1}\otimes B_k.
\end{equation}
Suppose that the channels commute under composition,
\begin{equation}
    \mathcal{F}\circ\mathcal{E}
    =
    \mathcal{E}\circ\mathcal{F}.
    \label{eq:channel-commutation}
\end{equation}
The two families are then Kraus representations of the same channel and contain the same number of operators. By the unitary freedom in the Kraus representations~\cite{Watrous2018}, there exists an
$N\times N$ unitary matrix $U$ such that
\begin{equation}
    B_k
    =
    \sum_{l=1}^{N}U_{kl}A_l.
    \label{eq:kraus-unitary-relation}
\end{equation}
Since $U$ is a unitary operator, there always exists another unitary operator $R$ diagonalizing it in the form 
\begin{equation}
    R U R^\dagger
    =
    \operatorname{diag}
    \left(
        e^{i\theta_1},
        \ldots,
        e^{i\theta_N}
    \right),
    \label{eq:kraus-unitary-diagonalization}
\end{equation}
for some $\theta_k\in\mathbb{R}$. We now define new Kraus families by employing the same unitary operation for both sets,
\begin{equation}
    A'_k
    =
    \sum_l R_{kl}A_l,
    \qquad
    B'_k
    =
    \sum_l R_{kl}B_l.
\end{equation}
Using Eqs.~\eqref{eq:kraus-unitary-relation} and~\eqref{eq:kraus-unitary-diagonalization}, we obtain
\begin{align}
    B'_k
    &=
    \sum_l\left(R U R^\dagger\right)_{kl}
    A'_l
    =
    e^{i\theta_k}A'_k.
\end{align}
Applying the same unitary change to the Kraus operators of the quantum switch gives
\begin{align}
    S'_k
    &=
    \sum_lR_{kl}S_l\nonumber\\
    &=
    \ket{0}\bra{0}\otimes A'_k
    +
    \ket{1}\bra{1}\otimes B'_k\nonumber\\
    &=
    \left(\ket{0}\bra{0}
        +
        e^{i\theta_k}\ket{1}\bra{1}\right)
    \otimes A'_k.
    \label{eq:factorized-switch-kraus}
\end{align}
Because $R$ is a unitary, the family $\{S'_k\}$ represents the same switched channel as the original Kraus family $\{S_k\}$. Equation~\eqref{eq:factorized-switch-kraus} shows that every
$S'_k$ is a product operator across the control-target partition, which means that starting from a separable input, it is impossible to generate entanglement switching two quantum channels that commute under composition. We stress that Eq.~\eqref{eq:channel-commutation} is a condition on the
channels as maps and does not require pairwise commutation of their Kraus operators. 

This condition also reproduces the input-independent nonentangling conditions previously found in the unitary analysis. For unitary channels, commutation as maps means $\mathcal{V}\circ\mathcal{U}=\mathcal{U}\circ\mathcal{V}$, which is equivalent to $VU=e^{i\chi}UV$
for some $\chi\in\mathbb{R}$. For qubit unitaries, taking determinants of both sides gives $e^{2i\chi}=1$, and hence $e^{i\chi}=\pm1$. Channel commutation clearly includes both
\begin{equation}
    [U,V]=0
    \qquad\text{and}\qquad
    \{U,V\}=0,
\end{equation}
which are exactly the same two cases corresponding to $X=V^\dagger U^\dagger VU=\pm\mathbb{I}$ identified in Sec.~\ref{sec:unitary-switch}. The general result also includes the self-switch, since every channel commutes with itself, and the switch of two Pauli channels, since Pauli channels commute under composition. It also accounts for the vanishing of the concurrence along $p_x=p_z$ in Fig.~\ref{fig3}(a), since the Hadamard and Pauli channels commute as channels precisely under this condition. 

Consequently, whenever the switch generates nonzero control-target entanglement from a separable input, the two ordered compositions must be different,
\begin{equation}
    \mathcal{F}\circ\mathcal{E}
    \neq
    \mathcal{E}\circ\mathcal{F}.
\end{equation}
Hence, they define two distinct quantum channels, which means that there exists at least one target input for which the two orders produce different output states. We note that entanglement generation witnesses the inequivalence of the two alternatively ordered compositions, although it does not by itself establish causal nonseparability. However, noncommutation is not sufficient to guarantee entanglement for every separable input. As demonstrated below, some noncommuting channel pairs produce a separable output for certain input states.

\subsection{Dissipative Channels}

\begin{figure}[t]
\centering
\includegraphics[width=\columnwidth]{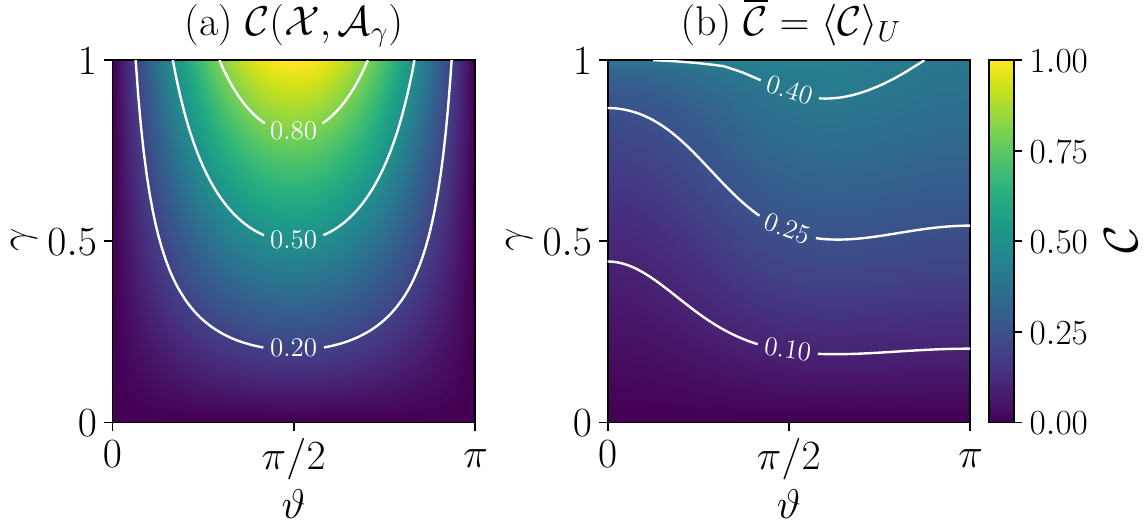}
\caption{Density plots of the concurrence generated by the switch of the
amplitude damping $\mathcal{A}_{\gamma}$ and unitary channels, for the maximally coherent control state and the real target input
    $\ket{\psi(\vartheta)}
    =
    \cos(\vartheta/2)\ket{0}
    +
    \sin(\vartheta/2)\ket{1}$.
(a) Concurrence for the quantum switch of bit-flip and amplitude damping channels. (b) Haar-averaged concurrence for the switch of unitary and amplitude damping channels, obtained by averaging over $U$ while keeping the control and target fixed.}
\label{fig4}
\end{figure}

The noisy channels considered so far are all unital, and their Kraus operators are proportional to unitaries. We now relax this restriction and take into account dissipative processes. Their Kraus operators are no longer proportional to unitaries, and instead of preserving the maximally mixed state, they contract the Bloch sphere toward a fixed point. We first consider the quantum switch of the zero-temperature amplitude damping channel $\mathcal{A}_{\gamma}$ and the unitary bit-flip channel
$\mathcal{X}(\rho)=\sigma_x\rho\sigma_x$. The amplitude damping channel is described by
$\mathcal{A}_{\gamma}(\rho)=\sum_{i=0}^{1}A_i\rho A_i^\dagger$, with the corresponding Kraus operators given by
\begin{align}
    A_0
    &=
    \ket{0}\bra{0}
    +
    \sqrt{1-\gamma}\ket{1}\bra{1},
    \nonumber\\
    A_1
    &=
    \sqrt{\gamma}\ket{0}\bra{1}.
    \label{eq:AD-kraus}
\end{align}
Here, $0\leq\gamma\leq1$ is the damping parameter. The bit-flip is a natural and analytically tractable choice here, since the amplitude damping process drives the target toward the ground state, whereas $\sigma_x$ maps that state to the excited one. Taking $\mathcal{E}=\mathcal{X}$ and $\mathcal{F}=\mathcal{A}_{\gamma}$ in
Eq.~\eqref{eq:switch-kraus}, the switched Kraus operators take the form
\begin{equation}
    S_i
    =
    \ket{0}\bra{0}\otimes A_i\sigma_x
    +
    \ket{1}\bra{1}\otimes\sigma_x A_i,
    \qquad
    i=0,1.
\end{equation}
We consider the control state in Eq.~\eqref{eq:control-input} and parametrize an arbitrary pure target state as
\begin{equation}
    \ket{\psi}
    =
    \cos(\vartheta/2)\ket{0}
    +
    e^{i\varphi}\sin(\vartheta/2)\ket{1},
    \label{eq:AD-X-target}
\end{equation}
where $0\leq\vartheta\leq\pi$ and $0\leq\varphi<2\pi$. The switched output has rank at most two and can be written as
\begin{equation}
    \rho_{CT}
    =
    \ket{w_0}\bra{w_0}
    +
    \ket{w_1}\bra{w_1},
\end{equation}
where the two unnormalized vectors are
\begin{equation}
    \ket{w_i}
    =
    \alpha\ket{0}\otimes
    A_i\sigma_x\ket{\psi}
    +
    \beta\ket{1}\otimes
    \sigma_xA_i\ket{\psi},
    \quad
    i=0,1.
    \label{eq:AD-X-vectors}
\end{equation}
Constructing the spin-flip overlap matrix introduced in Appendix~\ref{appB}, the concurrence is calculated as
\begin{equation}
    \mathcal{C}(\rho_{CT})
    =
    \sqrt{\gamma}\,|\alpha\beta|\,
    \left|\Lambda_+-\Lambda_-\right|,
    \label{eq:AD-X-angular-concurrence}
\end{equation}
where
\begin{equation}
\Lambda_{\pm} = \sqrt{(\cos \varphi \pm \sqrt{\gamma} \sin \vartheta)^2 + \cos^2 \vartheta \sin^2 \varphi}.
    \label{eq:AD-X-Lambda}
\end{equation}
Eqs.~\eqref{eq:AD-X-angular-concurrence} and~\eqref{eq:AD-X-Lambda} are derived in
Appendix~\ref{appD}. In particular, the control state enters the concurrence only through the factor $|\alpha\beta|$. As a result, for every fixed target input and damping parameter, the concurrence is maximized by a maximally coherent control. As $\Lambda_\pm\geq0$, they coincide if and only if their squares do, and
Eq.~\eqref{eq:AD-X-Lambda} gives
\begin{equation}
    \Lambda_+^2-\Lambda_-^2
    =
    4\sqrt{\gamma}\,\sin\vartheta\cos\varphi .
\end{equation}
Thus, for $\gamma>0$ and $\alpha\beta\neq0$, entanglement is generated if and only if $\sin\vartheta\cos\varphi\neq 0$. Since $\bra{\psi}\sigma_x\ket{\psi}= \sin\vartheta\cos\varphi$, the target Bloch vector must have a nonzero component along the $x$ axis selected by the bit-flip. This condition is not determined by the populations in the damping basis alone. For example, the
target state $\ket{+}=(\ket{0}+\ket{1})/\sqrt{2}$ generates entanglement, while
$(\ket{0}+i\ket{1})/\sqrt{2}$ does not. These findings are clearly illustrated in Fig.~\ref{fig4} for a maximally coherent control and real target inputs. Fig.~\ref{fig4}~(a) shows that, for the bit-flip channel, the concurrence vanishes at $\vartheta=0,\pi$ and increases with $\gamma$, reaching its largest value for the equatorial target $\vartheta=\pi/2$. Fig.~\ref{fig4}~(b) shows the corresponding Haar average over general unitary operators $U$. The Haar-averaged concurrence is nonzero for $\gamma>0$ even at $\vartheta=0,\pi$, demonstrating that entanglement generation is not restricted to the bit-flip channel.

Moreover, the concurrence can be optimized analytically over the control and the target inputs. As shown in Appendix~\ref{appD},
Eq.~\eqref{eq:AD-X-angular-concurrence} obeys
\begin{equation}
    \mathcal{C}(\rho_{CT})
    \leq
    2\gamma|\alpha\beta|\sin\vartheta
    \leq
    \gamma.
    \label{eq:AD-X-upper-bound}
\end{equation}
For a fixed $\vartheta$, the first inequality is saturated when $\varphi=0$ or $\pi$, corresponding to real amplitudes. In
these cases,
\begin{equation}
    \mathcal{C}(\rho_{CT})
    =
    2\gamma|\alpha\beta|\sin\vartheta.
    \label{eq:AD-X-real-concurrence}
\end{equation}
Hence, a complex relative phase cannot increase the concurrence, and the optimization can be restricted to real target states without loss of generality. The second inequality in Eq.~\eqref{eq:AD-X-upper-bound} is saturated for a maximally coherent control and an equatorial target state, $\vartheta=\pi/2$. As a result, both inequalities become equalities for a maximally coherent control state and $\ket{\psi}=\ket{+}$, yielding
\begin{equation}
    \mathcal{C}_{\max}
    \left(\mathcal{X},\mathcal{A}_{\gamma}\right)
    =
    \gamma,
    \label{eq:AD-X-maximum}
\end{equation}
where the maximization is over the control and target inputs. At $\gamma=0$, the amplitude damping channel reduces to the identity channel and no entanglement can be generated, in accordance with Eq.~\eqref{eq:AD-X-angular-concurrence}. On the other hand, at complete damping, $\gamma=1$, the Kraus operators reduce to $A_0=\ket{0}\!\bra{0}$ and $A_1=\ket{0}\!\bra{1}$ and the amplitude damping channel takes the form
\begin{equation}
    \mathcal{A}_{\gamma=1}(\rho)
    =
    \ket{0}\!\bra{0}\mathrm{Tr}(\rho)
    = \ket{0}\!\bra{0}.
\end{equation}
Indeed, complete amplitude damping is a replacer channel, which outputs the fixed ground state independently of the input, meaning that it is entanglement-breaking. Nevertheless, for the product input $\ket{+}_C\ket{+}_T$, its switch with the bit-flip channel outputs a pure two-qubit state, namely, the maximally entangled Bell state
\begin{equation}
    \ket{\Phi^+}
    =
    (1/\sqrt{2})(\ket{00}+\ket{11}).
\end{equation}
This result remarkably implies that although one of the inserted channels is entanglement-breaking, the switched process can still generate maximal control–target entanglement from a product input.

Next, we consider the switch of two zero-temperature amplitude damping channels, $\mathcal{A}_{\gamma_1}$ and $\mathcal{A}_{\gamma_2}$, which relax toward the same state with potentially different damping parameters. A direct composition gives
\begin{equation}
    \mathcal{A}_{\gamma_2}
    \circ
    \mathcal{A}_{\gamma_1}
    =
    \mathcal{A}_{\gamma_1+\gamma_2-\gamma_1\gamma_2}
    =
    \mathcal{A}_{\gamma_1}
    \circ
    \mathcal{A}_{\gamma_2}.
\end{equation}
Thus, the two amplitude damping channels commute under composition for arbitrary $\gamma_1,\gamma_2\in[0,1]$, including the case $\gamma_1\neq\gamma_2$. The result established in Sec.~\ref{sec:channel-commutation} implies that their switch maps every separable control-target input to a separable output state. In particular, for the pure product inputs considered throughout this work,
\begin{equation}
    \mathcal{C}(\rho_{CT})=0,
    \qquad
    \mathcal{C}_{\max}
    \left(
        \mathcal{A}_{\gamma_1},
        \mathcal{A}_{\gamma_2}
    \right)
    =0.
\end{equation}
This means that unequal damping strengths alone cannot generate control-target entanglement through the switch. However, this result still leaves open whether entanglement can be generated when two dissipative channels relax toward different equilibrium states. This is exactly the question that we intend to address next by extending our analysis of the switched dissipative channels to finite-temperature relaxation channels.

Let us consider the switch of two generalized amplitude damping channels, denoted by $\mathcal{G}_{\gamma_k,p_k}$ with $k\in\{1,2\}$. Here, $k$ labels the two channels entering the switch. For each channel, $\gamma_k\in[0,1]$ denotes the damping parameter, while $p_k\in[0,1]$ gives the stationary population of $\ket{0}$. Each quantum channel is described by a set of four Kraus operators $G_i^{(k)}$, with $i\in\{0,1,2,3\}$, as
\begin{equation}
    \mathcal{G}_{\gamma_k,p_k}(\rho)
    =
    \sum_{i=0}^{3}
    G_i^{(k)}\rho G_i^{(k)\dagger},
    \qquad
    k\in\{1,2\},
\end{equation}
where the operators can be written as
\begin{align}
    G_0^{(k)}
    &=
    \sqrt{p_k}
    \left(
        \ket{0}\bra{0}
        +
        r_k\ket{1}\bra{1}
    \right),
    \nonumber\\
    G_1^{(k)}
    &=
    \sqrt{p_k\gamma_k}\,
    \ket{0}\bra{1},
    \nonumber\\
    G_2^{(k)}
    &=
    \sqrt{1-p_k}
    \left(
        r_k\ket{0}\bra{0}
        +
        \ket{1}\bra{1}
    \right),
    \nonumber\\
    G_3^{(k)}
    &=
    \sqrt{(1-p_k)\gamma_k}\,
    \ket{1}\bra{0},
    \label{eq:GAD-kraus}
\end{align}
with $r_k=\sqrt{1-\gamma_k}$. For $\gamma_k>0$, this damping channel has the steady state described by the density operator
\begin{equation}
    \rho_k^{ss}
    =
    p_k\ket{0}\bra{0}
    +
    (1-p_k)\ket{1}\bra{1},
\end{equation}
which indicates that while $p_k=0$ corresponds to the inverse amplitude damping channel, $p_k=1$ gives the zero-temperature amplitude damping channel. When the channels have the same stationary state, i.e.,
$p_1=p_2=p$, their composition satisfies
\begin{equation}
    \mathcal{G}_{\gamma_2,p}
    \circ
    \mathcal{G}_{\gamma_1,p}
    =
    \mathcal{G}_{\gamma_1+\gamma_2-\gamma_1\gamma_2,p}
    =
    \mathcal{G}_{\gamma_1,p}
    \circ
    \mathcal{G}_{\gamma_2,p}.
    \label{eq:GAD-common-p-composition}
\end{equation}
It is clear that the channels commute under composition for arbitrary $\gamma_1$ and $\gamma_2$. The result of Sec.~\ref{sec:channel-commutation} immediately implies that their switch cannot generate control-target entanglement from a separable input. This shows that different damping parameters alone are insufficient when the two channels relax toward the same stationary state. In fact, the two generalized amplitude damping channels commute under composition when
\begin{equation}
    \gamma_1\gamma_2(p_1-p_2)=0,
    \label{eq:GAD-commutation-condition}
\end{equation}
as shown in Appendix~\ref{appE}. Consequently, when both damping parameters are nonzero, different stationary populations make the channels noncommuting under composition. Then, the separability conclusion that is established for commuting channels in Sec.~\ref{sec:channel-commutation} no longer follows, even though noncommutation alone does not guarantee entanglement generation for a given input state. For fixed channel parameters, let
\begin{equation}
    \mathcal{C}_{\max}
    \left(\mathcal{G}_{\gamma_1,p_1},\mathcal{G}_{\gamma_2,p_2}\right)
    =
    \max_{\ket{\psi_C},\ket{\psi}}\mathcal{C}(\rho_{CT}),
\end{equation}
where the maximization is over pure control and target inputs. As demonstrated in Appendix~\ref{appE}, the maximum concurrence obeys the inequality
\begin{equation}
    \mathcal{C}_{\max}
    \left(\mathcal{G}_{\gamma_1,p_1},\mathcal{G}_{\gamma_2,p_2}\right)
    \leq
    \frac{\gamma_1\gamma_2}{2}|p_1-p_2|.
    \label{eq:GAD-concurrence-bound}
\end{equation}
The bound clearly shows that different stationary populations and nonzero damping in both channels are necessary for the generation of entanglement, whereas different damping parameters alone are insufficient. This bound is saturated by a simple family of channels and inputs. For instance, considering the complete damping case with
\begin{equation}
    \gamma_1=\gamma_2=1,
    \qquad
    p_1=1,
    \qquad
    p_2=p,
\end{equation}
and preparing both the control and target in $\ket{+}$ gives
\begin{equation}
    \mathcal{C}(\rho_{CT})
    = (1-p)/2
    \label{eq:GAD-saturating-family}
\end{equation}
which saturates Eq.~\eqref{eq:GAD-concurrence-bound}. In particular, over the complete parameter range $0\leq p_k\leq1$, the global maximum is $\mathcal{C}_{\max}= 1/2$, which can be
attained for complete damping, opposite pure stationary states $(p_1,p_2)=(1,0)$ or $(0,1)$, and maximally coherent control and target inputs. If $\ket{0}$ is the ground state of a fixed qubit Hamiltonian and the stationary states are restricted to Gibbs states at nonnegative temperature, for which $1/2\leq p_k\leq1$, the corresponding maximum is instead $\mathcal{C}_{\max}=1/4$. The generalized amplitude damping family provides a direct application of the channel commutation result. Unequal damping strengths do not generate entanglement when the stationary state is common, while different stationary populations can generate control-target entanglement.

\begin{figure}[t]
\centering
\includegraphics[width=\columnwidth]{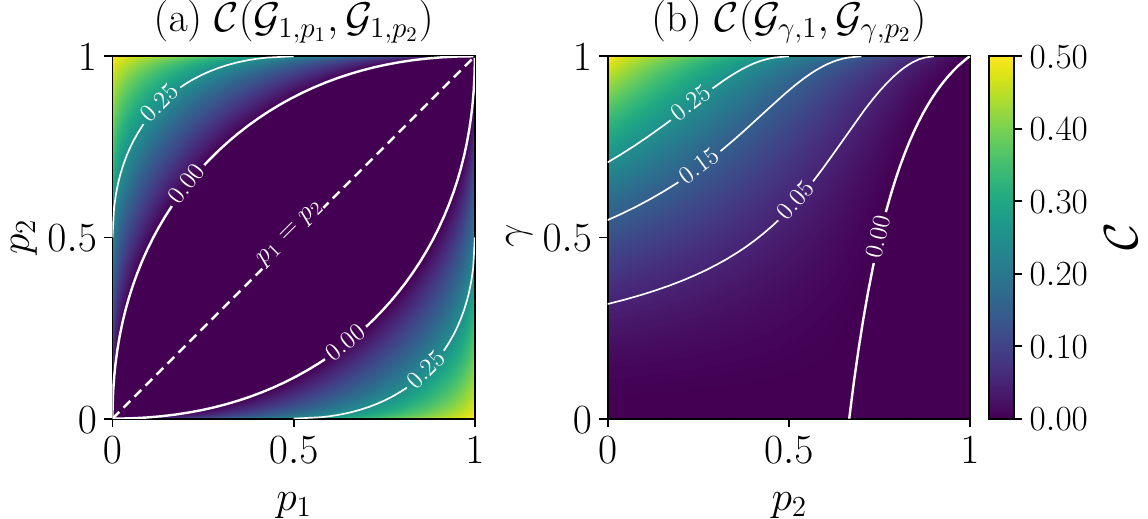}
\caption{Density plots of the concurrence generated by the switch of two generalized amplitude damping channels, $\mathcal{G}_{\gamma_1,p_1}$ and $\mathcal{G}_{\gamma_2,p_2}$, for the product input $\ket{+}_C\ket{+}_T$. (a) Both channels are fully relaxing, i.e., $\gamma_1=\gamma_2=1$. The dashed diagonal marks $p_1=p_2$, for which the channels commute under composition and the switched output is separable. (b) The first stationary population is fixed at $p_1=1$, while the channels have a common damping strength, $\gamma_1=\gamma_2=\gamma$. The thicker solid white curves are the  $\mathcal{C}=0$ boundaries.}
\label{fig5}
\end{figure}

Figure~\ref{fig5} illustrates these results for the fixed product input $\ket{+}_C\ket{+}_T$. For complete relaxation, Fig.~\ref{fig5}(a) displays a finite zero concurrence region for the chosen input that contains the commuting line $p_1=p_2$. Away from this region, the concurrence generally becomes larger toward parameter values corresponding to strongly different stationary populations. At $(p_1,p_2)=(1,0)$ and $(0,1)$, it reaches $\mathcal{C}=1/2$, which is the largest value permitted by Eq.~\eqref{eq:GAD-concurrence-bound}. Figure~\ref{fig5}(b) shows how the entangled region develops with increasing common damping parameter when $p_1=1$. The concurrence reaches the same maximum at $(p_2,\gamma)=(0,1)$. The zero concurrence region extends into parameter values for which the two channels do not commute, demonstrating that channel noncommutation alone does not guarantee an entangled output for the displayed input. Moreover, nonzero concurrence is obtained for $p_2>1/2$ at sufficiently strong damping. Hence, population inversion, i.e., negative temperatures, is not required to generate control-target entanglement.

\section{Entangling capability of the quantum time-flip}
\label{sec4}

We now turn from coherent control of channel order to coherent control of input-output direction. We first focus on its entangling capability for an arbitrary unitary qubit channel and then its simplest noisy extension, namely, a binary random unitary channel.

Before considering particular channels, we briefly discuss a general condition that prevents entanglement generation by the quantum time-flip. In the fixed basis defining the transposition, let
\begin{equation}
    \mathcal{E}^{ T}(\rho)
    =
    \sum_i E_i^{ T}\rho E_i^\ast
\end{equation}
denote the channel corresponding to the reversed input-output direction. It has been shown that a transposition invariant channel, $\mathcal{E}^{ T}=\mathcal{E}$, admits a Kraus representation
$\{N_i\}$ in which every operator is either symmetric or antisymmetric under transposition~\cite{Liu2023b},
\begin{equation}
    N_i^{T}
    =
    \eta_iN_i,
    \qquad
    \eta_i=\pm1.
\end{equation}
Then, the corresponding Kraus operators  factorize as
\begin{align}
    T_i
    &=
    \ket{0}\bra{0}\otimes N_i
    +
    \ket{1}\bra{1}\otimes N_i^{T}\nonumber\\
    &=
    \left(\ket{0}\bra{0}
        +
    \eta_i\ket{1}\bra{1}\right)
    \otimes N_i,
\end{align}
which implies that every $T_i$ is a product operator across the control-target partition, and the time-flip maps every separable control-target input to a separable output. Transposition invariance provides a general separability condition for the quantum time-flip, analogous to channel commutation for the quantum switch.

\subsection{Unitary Channels}

Let us first consider the quantum time-flip of a unitary qubit channel, $\mathcal{U}(\rho)=U\rho U^\dagger$. The corresponding time-flip operator is given by
\begin{equation}
    T_U
    =
    \ket{0}\bra{0}\otimes U
    +
    \ket{1}\bra{1}\otimes U^{T},
\end{equation}
where the computational basis of the control qubit labels the forward and backward input-output directions of the channel. For the control state given in Eq.~\eqref{eq:control-input}, the target is prepared in an arbitrary pure state $\ket{\psi}$. Acting with $T_U$ on the product input
$\ket{\psi_C}\otimes\ket{\psi}$ yields
\begin{align}
    \ket{\Phi}
    &=
    \alpha\ket{0}\otimes U\ket{\psi}
    +
    \beta\ket{1}\otimes U^{T}\ket{\psi}
    \nonumber\\
    &=
    \alpha\ket{0}\otimes\ket{\phi_0}
    +
    \beta\ket{1}\otimes\ket{\phi_1},
    \label{eq:unitary-time-flip-output}
\end{align}
where we introduced $\ket{\phi_0}=U\ket{\psi}$ and $\ket{\phi_1}=U^{T}\ket{\psi}$, denoting the forward and backward evolutions of the target. The overlap of these two states,
\begin{equation}
    \omega
    =
    \braket{\phi_1|\phi_0}
    =
    \bra{\psi}Y\ket{\psi},
    \qquad
    Y=U^\ast U,
    \label{eq:time-flip-overlap}
\end{equation}
quantifies how distinguishable the two directional evolutions are. Since $Y$ is a product of unitary operators, it is itself unitary, so we have $|\omega|\leq1$ for all target states $\ket{\psi}$. The state in Eq.~\eqref{eq:unitary-time-flip-output} is a pure two-qubit state, and it is separable if and only if either $\alpha\beta=0$ or $\ket{\phi_0}$ and $\ket{\phi_1}$ coincide up to a phase, equivalently $|\omega|=1$. Entanglement is generated when $\alpha\beta\neq0$ and $|\omega|<1$. Utilizing the pure two-branch result in Eq.~\eqref{eq:concurrence-s}, the concurrence of the output state $\varrho_{CT}=\ket{\Phi}\bra{\Phi}$ becomes
\begin{equation}
    \mathcal{C}(\varrho_{CT})
    =
    2|\alpha\beta|\sqrt{1-|\omega|^2},
    \label{eq:unitary-time-flip-concurrence}
\end{equation}
which has the same form as the switch result in Eq.~\eqref{eq:concurrence-s}. Since the time-flip output has the same  structure as the unitary switch output, the coherence-entanglement conservation relation in Eq.~\eqref{eq:coherence-concurrence-l1} also holds here, that is,
\begin{equation}
    C_{\ell_1}(\rho_C^{\mathrm{in}})^2
    =
    C_{\ell_1}(\varrho_C)^2
    +
    \mathcal{C}(\varrho_{CT})^2.
\end{equation}
As a result, the squared initial control coherence is again divided between the squared residual control coherence and the squared concurrence of the control and target.

For fixed $U$ and $\ket{\psi}$, the concurrence is maximized by a balanced control state, $|\alpha|=|\beta|=1/\sqrt{2}$. The remaining optimization is most naturally expressed through the
spectrum of $Y$. Due to the fact that $\det Y=1$, we have $Y\in SU(2)$, and its eigenvalues form a conjugate pair,
\begin{equation}
    Y\ket{\xi_\pm}
    =
    e^{\pm i\theta_Y/2}\ket{\xi_\pm},
    \qquad
    0\leq\theta_Y\leq2\pi.
    \label{eq:time-flip-spectrum}
\end{equation}
As shown in Appendix~\ref{appF}, for $Y\neq\pm\mathbb{I}$ the magnitude of the overlap is minimized by an equal weight superposition of $\ket{\xi_+}$ and $\ket{\xi_-}$. Maximizing also over the control gives
\begin{equation}
    \begin{aligned}
        \mathcal{C}_{\max}(U)
        &\equiv
        \max_{\ket{\psi_C},\,\ket{\psi}}
        \mathcal{C}(\varrho_{CT})
        \\
        &=
        \left|\sin(\theta_Y/2)\right|
        =
        \sqrt{1-(1/4)\left(\mathrm{Tr}\,Y\right)^2}.
    \end{aligned}
    \label{eq:unitary-time-flip-spectral-maximum}
\end{equation}
If $Y=\pm\mathbb{I}$, every target state gives zero concurrence. At the opposite extreme, the time-flip can generate a maximally entangled state if and only if $\theta_Y=\pi$, equivalently $\mathrm{Tr}\,Y=0$, so that the eigenvalues of $Y$ are $\{i,-i\}$. The maximal concurrence also admits a closed form expression in terms of the geometric parameters of $U$. Since the global phase of $U$ cancels in $Y$, we may take $U\in SU(2)$ without loss of generality and write
\begin{equation}
    U
    =
    \cos(\theta_U/2)\,\mathbb{I}
    -
    i\sin(\theta_U/2)\,
    (\hat{n}\cdot\vec{\sigma}),
    \label{eq:time-flip-axisangle}
\end{equation}
where  $\hat{n}=(n_x,n_y,n_z)$ is a unit vector along the rotation axis and $\theta_U\in[0,2\pi]$ is the rotation angle. Transposition reverses only the $y$ component of this axis, and a direct
evaluation gives (see Appendix~\ref{appF})
\begin{equation}
    \mathrm{Tr}\,Y/2
    =
    \cos(\theta_Y/2)
    =
    1-2n_y^2\sin^2(\theta_U/2).
    \label{eq:time-flip-trace-axisangle}
\end{equation}
Defining $\kappa=n_y^2\sin^2(\theta_U/2)$, with $0\leq\kappa\leq1$, Eqs.~\eqref{eq:unitary-time-flip-spectral-maximum} and~\eqref{eq:time-flip-trace-axisangle} together give
\begin{equation}
    \mathcal{C}_{\max}(U)
    =
    2\sqrt{\kappa(1-\kappa)}.
    \label{eq:unitary-time-flip-maximum}
\end{equation}
The time-flip is maximally entangling provided $\kappa=1/2$, which requires $|n_y|\geq1/\sqrt{2}$ and $\pi/2\leq\theta_U\leq3\pi/2$. 

\subsection{Binary Random Unitary Channels}

We consider the simplest noisy extension of the unitary setting, namely, the binary random unitary channel
\begin{equation}
    \mathcal{R}_{p,U}(\rho)
    =
    (1-p)\rho
    +
    pU\rho U^\dagger,
    \qquad
    0\leq p\leq1,
\end{equation}
where $p$ is the noise parameter. A Kraus representation of this map is given by $K_0=\sqrt{1-p}\,\mathbb{I}$ and $K_1=\sqrt{p}\,U$. Under the quantum time-flip, the identity component leaves the product input unchanged, whereas the second component produces the unitary time-flip state $\ket{\Phi}$ considered above. The resulting output can be written as
\begin{equation}
    \begin{aligned}
        \varrho_{CT}
        =
        (1-p)\rho_C^{\mathrm{in}}
        \otimes
        \ket{\psi}\bra{\psi}
        +
        p\ket{\Phi}\bra{\Phi}.
    \end{aligned}
\end{equation}
Even though this state is generally mixed, its decomposition has the same two-term form treated in Appendix~\ref{appB}, that is, a rank-one product contribution and a rank-one unitary contribution. Therefore, the spin-flip overlap calculation leading to Eq.~\eqref{eq:concurrence-M11-appB} applies unchanged here, with $\ket{\Psi}$ replaced by $\ket{\Phi}$. Consequently, we obtain
\begin{equation}
    \begin{aligned}
        \mathcal{C}\bigl(\varrho_{CT}\bigr)
        =
        p\,
        \mathcal{C}\bigl(\ket{\Phi}\bra{\Phi}\bigr)
        =
        2p|\alpha\beta|
        \sqrt{1-|\omega|^2}.
    \end{aligned}
\end{equation}
Thus, for every $p>0$, entanglement is generated under the same conditions as in the unitary case. The maximizing control and target states are unchanged, and
\begin{equation}
    \begin{aligned}
        \mathcal{C}_{\max}
        \bigl(\mathcal{R}_{p,U}\bigr)
        =
        p\left|\sin(\theta_Y/2)\right|
        =
        2p\sqrt{\kappa(1-\kappa)}.
    \end{aligned}
\end{equation}
For a fixed value of $p$, the largest attainable concurrence is $p$, reached when $\kappa=1/2$. Hence, the quantum time-flip can generate entanglement throughout the noisy regime $0<p<1$, whereas a maximally entangled output state requires both $p=1$ and $\kappa=1/2$.

\section{Conclusion and Discussion}
\label{sec5}

We have presented a systematic analytical treatment of how coherent control of channel order and input-output direction can generate entanglement between control and target qubits. For the quantum switch of two arbitrary qubit unitary channels, we obtained the concurrence for arbitrary pure control and target inputs and determined its maximum over both inputs. We then expressed these results geometrically in terms of the rotation angles and axes of the two unitaries, which allowed us to identify both the maximally entangling pairs and those that cannot generate entanglement from any pure product input state. We also obtained an exact coherence-entanglement conservation relation showing how the coherence initially stored in the control system is divided between the remaining local coherence and the entanglement generated with the target. Due to the fact that the initial control coherence is fixed by preparation, this simple relation allows the generated concurrence to be obtained from measurements performed solely on the control qubit, without reconstructing the full control-target output state. 

Beyond the unitary setting, we extended our analysis to several canonical noise channels. For the switch of a binary random unitary channel and a unitary channel, we demonstrated that the concurrence and its maximum retain the same dependence on the underlying unitaries as in the purely unitary case, reduced by a simple factor determined by the noise parameter. More importantly, we established that whenever two channels commute under composition, their quantum switch cannot generate entanglement from any separable input state. This condition explains why self-switches, switches of two Pauli channels, switches of two zero-temperature amplitude damping channels, and switches of two generalized amplitude damping channels with the same stationary state are all nonentangling. However, we also showed that neither noise nor dissipation rules out entanglement generation. The clearest example is complete amplitude damping. Even though it is an entanglement-breaking replacer channel that erases all information about its input when acting alone, its quantum switch with a bit-flip can map a product input into a maximally entangled Bell state. This shows that the entangling capability belongs to the coherently controlled process as a whole and cannot be inferred from either inserted channel in isolation. Finally, for switches of two generalized amplitude damping channels, we showed that different stationary populations can enable entanglement. We then derived an analytical upper bound on the concurrence and identified a complete relaxation family that saturates it, including the parameter choices that attain the global maximum.

Finally, we considered the quantum time-flip as a complementary setting in which the input-output direction of a channel is coherently controlled. For unitary channels, we derived the concurrence and its maximum, and characterized when the two alternative directions can generate entanglement. We extended this analysis to binary random unitary channels and showed that the noise parameter simply rescales the unitary concurrence, while leaving the conditions for entanglement generation unchanged. All in all, our findings provide a comprehensive account of the mechanisms that enable or prevent entanglement generation from separable states through coherent control of channel order or input-output direction.

\begin{acknowledgments}
G. K. is supported by the Scientific and Technological Research Council of Türkiye (TUBITAK) through the 100th Anniversary Incentive Award. During the preparation of this work, the authors used OpenAI GPT-5.6 Sol to assist with identifying the relevant literature, polishing analytical derivations, developing numerical code and improving the presentation. The authors independently checked and verified all incorporated suggestions and retain full responsibility for the content of this manuscript.
\end{acknowledgments}

\appendix

\section{Concurrence of the pure switch output in the axis-angle form}
\label{appA}

Starting from the axis-angle representations given in Eq.~\eqref{eq:UV-axisangle}, where
$\vec{\sigma}=(\sigma_x,\sigma_y,\sigma_z)$ denotes the vector of Pauli operators, let us introduce the shorthand notation
\begin{equation}
    c_U=\cos(\theta_U/2),\qquad
    s_U=\sin(\theta_U/2),
\end{equation}
and analogously \(c_V=\cos(\theta_V/2)\) and
\(s_V=\sin(\theta_V/2)\). The two unitaries and their adjoints can be written as
\begin{align}
    U&=c_U\mathbb{I}-is_U(\hat{a}\cdot\vec{\sigma}),
    &
    U^\dagger&=c_U\mathbb{I}+is_U(\hat{a}\cdot\vec{\sigma}),
    \nonumber\\
    V&=c_V\mathbb{I}-is_V(\hat{b}\cdot\vec{\sigma}),
    &
    V^\dagger&=c_V\mathbb{I}+is_V(\hat{b}\cdot\vec{\sigma}).
\end{align}
In the following, we repeatedly use the Pauli identity
\begin{equation}
    (\vec{x}\cdot\vec{\sigma})
    (\vec{y}\cdot\vec{\sigma})
    =
    (\vec{x}\cdot\vec{y})\mathbb{I}
    +
    i(\vec{x}\times\vec{y})\cdot\vec{\sigma}.
    \label{eq:pauli-product-app}
\end{equation}
For convenience, let
\begin{align}
    A
    &=
    c_Vc_U-s_Vs_U(\hat{b}\cdot\hat{a}),
    \nonumber\\
    \vec{B}
    &=
    c_Vs_U\hat{a}+c_Us_V\hat{b},
    \nonumber\\
    \vec{C}
    &=
    s_Vs_U(\hat{b}\times\hat{a}).
\end{align}
Using Eq.~\eqref{eq:pauli-product-app}, the two product terms entering the operator
\(X=V^\dagger U^\dagger VU\) take the form
\begin{align}
    V^\dagger U^\dagger
    &=
    A\mathbb{I}
    +
    i(\vec{B}-\vec{C})\cdot\vec{\sigma},
    \nonumber\\
    VU
    &=
    A\mathbb{I}
    -
    i(\vec{B}+\vec{C})\cdot\vec{\sigma}.
\end{align}
Multiplication of these expressions gives
\begin{equation}
    X
    =
    \left(A^2+|\vec{B}|^2-|\vec{C}|^2\right)\mathbb{I}
    +
    2i\left(\vec{B}\times\vec{C}
    -A\vec{C}\right)\cdot\vec{\sigma}.
\end{equation}
Since every Pauli matrix is traceless, only the coefficient of the identity contributes to the trace. Therefore,
\begin{equation}
    \frac{\mathrm{Tr}\,X}{2}
    =
    A^2+|\vec{B}|^2-|\vec{C}|^2.
    \label{eq:trace-ABC-app}
\end{equation}
To evaluate this quantity, we set $d=\hat{a}\cdot\hat{b}=\cos\theta_{ab}$. The three terms in Eq.~\eqref{eq:trace-ABC-app} are then
\begin{align}
    A^2
    &=
    c_V^2c_U^2
    -2c_Vc_Us_Vs_Ud
    +s_V^2s_U^2d^2,
    \nonumber\\
    |\vec{B}|^2
    &=
    c_V^2s_U^2
    +c_U^2s_V^2
    +2c_Vs_Uc_Us_Vd,
    \nonumber\\
    |\vec{C}|^2
    &=
    s_V^2s_U^2|\hat{b}\times\hat{a}|^2
    =
    s_V^2s_U^2(1-d^2).
\end{align}
The terms linear in \(d\) cancel between \(A^2\) and \(|\vec{B}|^2\), giving
\begin{align}
    A^2+|\vec{B}|^2-|\vec{C}|^2
    &=
    c_V^2+s_V^2c_U^2+s_V^2s_U^2d^2
    \nonumber\\
    &-s_V^2s_U^2(1-d^2)
    \nonumber\\
    &=
    1-2s_V^2s_U^2(1-d^2).
\end{align}
Finally, using \(1-d^2=\sin^2\theta_{ab}\), we obtain
\begin{equation}
    \frac{\mathrm{Tr}\,X}{2}
    =
    1
    -
    2\sin^2(\theta_U/2)
     \sin^2(\theta_V/2)
     \sin^2(\theta_{ab}),
\end{equation}
which is Eq.~\eqref{eq:trace-axisangle}. Defining
\begin{equation}
    t=
    \sin^2(\theta_U/2)
    \sin^2(\theta_V/2)
    \sin^2(\theta_{ab}),
\end{equation}
we can write
\begin{equation}
    \cos(\theta_X/2)
    =
    \mathrm{Tr}\,X/2
    =
    1-2t.
\end{equation}
It then follows from Eq.~\eqref{eq:max-concurrence-spectral} that
\begin{align}
    \mathcal{C}_{\max}(U,V)
    &=
    \sqrt{1-\cos^2(\theta_X/2)}
    \nonumber\\
    &=
    \sqrt{1-(1-2t)^2}
    =
    2\sqrt{t(1-t)},
\end{align}
recovering Eq.~\eqref{eq:cmax-axisangle} in the main text.

\section{Concurrence of the mixed switch output}
\label{appB}

For the mixed two-qubit outputs considered below, it is convenient to evaluate the concurrence directly from a decomposition into $n$ unnormalized vectors,
\begin{equation}
    \rho
    =
    \sum_{i=0}^{n-1}
    \ket{w_i}\bra{w_i},
    \label{eq:rho-decomposition-appB}
\end{equation}
where $n$ denotes the number of vectors in the chosen decomposition. The corresponding spin-flipped vectors are defined as $\ket{\widetilde w_i} = (\sigma_y\otimes\sigma_y)\ket{w_i^\ast}$, where complex conjugation is taken in the computational basis. The spin-flip overlap matrix is then written as
\begin{equation}
    M_{ij}
    =
    \braket{w_i|\widetilde w_j},
    \qquad
    i,j=0,\ldots,n-1.
\end{equation}
Since
$(\sigma_y\otimes\sigma_y)^{ T}=\sigma_y\otimes\sigma_y$, it follows that $M_{ij}=M_{ji}$, and hence $M$ is symmetric. In fact, the nonzero singular values of $M$ coincide with the square roots of the nonzero eigenvalues of $\rho\widetilde{\rho}$, where $\widetilde{\rho}=  (\sigma_y\otimes\sigma_y)\rho^\ast (\sigma_y\otimes\sigma_y)$. They are determined by $\rho$ and are independent of the particular decomposition given in Eq.~\eqref{eq:rho-decomposition-appB}. Since $\rho\widetilde{\rho}$ acts on the four-dimensional two-qubit Hilbert
space, at most four of them are nonzero. The concurrence can be evaluated from these singular values~\cite{Wootters1998,Wootters2001} as
\begin{equation}
    \mathcal{C}(\rho)
    =
    \max\left\{
    0,\mu_1-\mu_2-\mu_3-\mu_4
    \right\},
    \label{eq:concurrence-from-M-appB}
\end{equation}
where the singular values are arranged in nonincreasing order, $\mu_1\geq\mu_2\geq\mu_3\geq\mu_4$, and zeros are appended when fewer than four are nonzero.

\subsection{Binary random unitary and unitary channels}

For the output state given in Eq.~\eqref{eq:binary-ru-output}, the decomposition contains two unnormalized vectors, $\ket{w_0}$ and $\ket{w_1}$.
The corresponding spin-flip overlap matrix is 
\begin{equation}
    M
    =
    \begin{pmatrix}
        M_{00} & M_{01}\\
        M_{01} & M_{11}
    \end{pmatrix}.
\end{equation}
Using Eq.~\eqref{eq:binary-ru-vectors}, its elements are
\begin{align}
    M_{00}
    &=
    (1-p)
    \braket{\chi|\widetilde{\chi}}
    =
    0,
    \nonumber\\
    M_{01}
    &=
    \sqrt{p(1-p)}
    \braket{\chi|\widetilde{\Psi}},
    \nonumber\\
    M_{11}
    &=
    p\braket{\Psi|\widetilde{\Psi}}.
\end{align}
Here, $M_{00}=0$ because $\ket{\chi}$ is a product state and has a vanishing overlap with its spin flip. Hence,
\begin{equation}
    M
    =
    \begin{pmatrix}
        0 & M_{01}\\
        M_{01} & M_{11}
    \end{pmatrix}.
\end{equation}
Let $\mu_1\geq\mu_2\geq0$ denote the two singular values of $M$. In Eq.~\eqref{eq:concurrence-from-M-appB}, the remaining singular values are $\mu_3=\mu_4=0$. The two singular values satisfy
\begin{align}
    \mu_1^2+\mu_2^2
    =
    \mathrm{Tr}\,(MM^\dagger)
    =
    2|M_{01}|^2+|M_{11}|^2,
\end{align}
while their product is
\begin{align}
    \mu_1\mu_2
    =
    \sqrt{\det(MM^\dagger)}
    =
    |\det M|
    =
    |M_{01}|^2.
\end{align}
It follows that
    $(\mu_1-\mu_2)^2
    =
    \mu_1^2+\mu_2^2
    -
    2\mu_1\mu_2
    =
    |M_{11}|^2.$
Since $\mu_1\geq\mu_2$, the concurrence becomes
\begin{align}
    \mathcal{C}(\rho_{CT})
    =
    \mu_1-\mu_2
    =
    |M_{11}|
    =
    p\,\lvert\braket{\Psi|\widetilde{\Psi}}\rvert.
    \label{eq:concurrence-M11-appB}
\end{align}
As the modulus of the spin-flip overlap equals its concurrence for a pure two-qubit state, we have
\begin{align}
    \mathcal{C}(\rho_{CT})
    =
    p\,\mathcal{C}
    \left(\ket{\Psi}\bra{\Psi}\right)
    =
    2p|\alpha\beta|
    \sqrt{1-|s|^2},
\end{align}
which proves Eq.~\eqref{eq:binary-ru-concurrence}. Although $M_{01}$ generally does not vanish, its contribution cancels in $\mu_1-\mu_2$ and does not appear in the final expression for the concurrence.

\subsection{Switch of two binary random unitary channels}

For the output state given in Eq.~\eqref{eq:two-binary-ru-output}, the decomposition contains four unnormalized vectors, $\ket{w_{00}}$, $\ket{w_{01}}$, $\ket{w_{10}}$, and $\ket{w_{11}}$. The spin-flip overlap matrix is 
\begin{equation}
    M_{ij,kl}
    =
    \braket{w_{ij}|\widetilde w_{kl}} .
\end{equation}
The first three vectors given in Eq.~\eqref{eq:two-binary-ru-vectors} are product states with the same control component. Therefore, their overlaps with their corresponding spin-flipped vectors vanish. Moreover, for any two of these vectors, the target states may differ, but the control part of the spin-flip overlap contains the factor $\langle\psi_C|\sigma_y|\psi_C^\ast\rangle$, which indeed vanishes for every single-qubit pure state. Then, all matrix elements for which both indices correspond to the first three vectors vanish and the overlap matrix has the form
\begin{equation}
M=
\begin{pmatrix}
0&0&0&M_{00,11}\\
0&0&0&M_{01,11}\\
0&0&0&M_{10,11}\\
M_{00,11}&M_{01,11}&M_{10,11}&M_{11,11}
\end{pmatrix},
\end{equation}
which has rank at most two. Diagonalization of the matrix $MM^\dagger$ gives the singular values of $M$ as
\begin{equation}
\begin{split}
    \mu_{1,2}
    ={}&\frac{1}{2}\Big[
        |M_{11,11}|^2
        +4|M_{00,11}|^2
        +4|M_{01,11}|^2
        \\&
        +4|M_{10,11}|^2
    \Big]^{1/2}
    \pm\frac{1}{2}|M_{11,11}|.
\end{split}
\end{equation}
Therefore, the concurrence can be calculated as
\begin{align}
    \mathcal{C}(\rho_{CT})
    =
    \mu_1-\mu_2
    =
    |M_{11,11}|
    =
    pq\,\lvert\braket{\Psi|\widetilde{\Psi}}\rvert.
\end{align}
Finally, using the pure state concurrence in Eq.~\eqref{eq:concurrence-s},
\begin{align}
    \mathcal{C}(\rho_{CT})
    =
    2pq|\alpha\beta|
    \sqrt{1-|s|^2},
\end{align}
which proves Eq.~\eqref{eq:two-binary-ru-concurrence-explicit}.

\section{Switch of Pauli and Hadamard channels}
\label{appC}

In this part, we first prove the upper bound in Eq.~\eqref{eq:pauli-H-upper-bound} for an arbitrary target input state and then show that it is saturated by a Hadamard eigenstate. Let us write the unnormalized vectors in Eq.~\eqref{eq:pauli-H-vectors} as
\begin{equation}
    \ket{w_\mu}
    =
    \sqrt{p_\mu}\ket{\Psi_\mu},
\end{equation}
where we introduce
\begin{equation}
\ket{\Psi_\mu}=\alpha\ket{0}\otimes\sigma_\mu H\ket{\psi} +\beta\ket{1}\otimes H\sigma_\mu\ket{\psi}.
\end{equation}
Each vector $\ket{\Psi_\mu}$ is normalized, and the output reads
\begin{equation}
    \rho_{CT}
    =
    \sum_{\mu=0,x,y,z}
    p_\mu
    \ket{\Psi_\mu}\bra{\Psi_\mu}.
    \label{eq:pauli-output-appC}
\end{equation}
Using that $H\sigma_x=\sigma_zH$, $H\sigma_z=\sigma_xH$, $H\sigma_y=-\sigma_yH$, the vectors associated with the identity and $\sigma_y$ become
\begin{align}
    \ket{\Psi_0}
    &=
    \left(
    \alpha\ket{0}
    +
    \beta\ket{1}
    \right)
    \otimes H\ket{\psi},
    \nonumber\\
    \ket{\Psi_y}
    &=
    \left(
    \alpha\ket{0}
    -
    \beta\ket{1}
    \right)
    \otimes\sigma_yH\ket{\psi}.
\end{align}
We note that both $\ket{\Psi_0}$ and $\ket{\Psi_y}$ are product states. The remaining two vectors take the form
\begin{align}
    \ket{\Psi_x}
    &=
    \alpha\ket{0}\otimes\sigma_xH\ket{\psi}
    +
    \beta\ket{1}\otimes\sigma_zH\ket{\psi},
    \nonumber\\
    \ket{\Psi_z}
    &=
    \alpha\ket{0}\otimes\sigma_zH\ket{\psi}
    +
    \beta\ket{1}\otimes\sigma_xH\ket{\psi}.
\end{align}
Their sum and difference factorize as
\begin{align}
    \ket{\Gamma_+}
    &=
    \ket{\Psi_x}
    +
    \ket{\Psi_z}
    \nonumber\\
    &=
    \left(
    \alpha\ket{0}
    +
    \beta\ket{1}
    \right)
    \otimes
    \left(
    \sigma_x+\sigma_z
    \right)
    H\ket{\psi},
    \nonumber\\
    \ket{\Gamma_-}
    &=
    \ket{\Psi_x}
    -
    \ket{\Psi_z}
    \nonumber\\
    &=
    \left(
    \alpha\ket{0}
    -
    \beta\ket{1}
    \right)
    \otimes
    \left(
    \sigma_x-\sigma_z
    \right)
    H\ket{\psi}.
\end{align}
Both $\ket{\Gamma_+}$ and $\ket{\Gamma_-}$ are in product form, implying that
\begin{align}
    \tau_{xz}
    &=
    \frac{1}{2}
    \left(
    \ket{\Psi_x}\bra{\Psi_x}
    +
    \ket{\Psi_z}\bra{\Psi_z}
    \right)
    \nonumber\\
    &=
    \frac{1}{4}
    \left(
    \ket{\Gamma_+}\bra{\Gamma_+}
    +
    \ket{\Gamma_-}\bra{\Gamma_-}
    \right)
    \label{eq:pauli-tau-xz-appC}
\end{align}
is separable. Suppose first that $p_x\geq p_z$. Using Eq.~\eqref{eq:pauli-tau-xz-appC}, the output state in Eq.~\eqref{eq:pauli-output-appC} can be rearranged as
\begin{align}
    \rho_{CT}
    ={}&
    p_0\ket{\Psi_0}\bra{\Psi_0}
    +
    p_y\ket{\Psi_y}\bra{\Psi_y}
    +
    2p_z\tau_{xz}
    \nonumber\\
    &+
    (p_x-p_z)
    \ket{\Psi_x}\bra{\Psi_x}.
\end{align}
The first three terms on the right-hand side form a separable contribution of total weight
$1-(p_x-p_z)$. Hence, using the convexity of concurrence, we obtain
\begin{equation}
    \mathcal{C}(\rho_{CT})
    \leq
    (p_x-p_z)
    \mathcal{C}
    \left(
    \ket{\Psi_x}\bra{\Psi_x}
    \right).
\end{equation}
As $\ket{\Psi_x}$ is a pure unitary switch output, Eq.~\eqref{eq:concurrence-s} implies
\begin{equation}
    \mathcal{C}
    \left(
    \ket{\Psi_x}\bra{\Psi_x}
    \right)
    \leq
    2|\alpha\beta|.
\end{equation}
It therefore follows that
\begin{equation}
    \mathcal{C}(\rho_{CT})
    \leq
    2|\alpha\beta|
    (p_x-p_z),
    \qquad
    p_x\geq p_z.
    \label{eq:pauli-upper-bound-x-appC}
\end{equation}
When $p_z\geq p_x$, the same argument gives
\begin{equation}
    \mathcal{C}(\rho_{CT})
    \leq
    2|\alpha\beta|
    (p_z-p_x).
    \label{eq:pauli-upper-bound-z-appC}
\end{equation}
Combining Eqs.~\eqref{eq:pauli-upper-bound-x-appC} and~\eqref{eq:pauli-upper-bound-z-appC}, we obtain
\begin{equation}
    \mathcal{C}(\rho_{CT})
    \leq
    2|\alpha\beta|
    |p_x-p_z|,
\end{equation}
which proves Eq.~\eqref{eq:pauli-H-upper-bound} in the main text.

As a last step, it remains to show that the upper bound above can be saturated. Indeed, it is sufficient to consider the Hadamard eigenstate
\begin{equation}
    \ket{h_+}
    =
    \cos\left(\pi/8\right)\ket{0}
    +
    \sin\left(\pi/8\right)\ket{1}.
\end{equation}
Utilizing the unnormalized vectors $\ket{w_\mu}$ given in Eq.~\eqref{eq:pauli-H-vectors}, we construct the spin-flip overlap matrix
\begin{equation}
    M_{\mu\nu}^{+}
    =
    \braket{w_\mu|\widetilde w_\nu},
    \qquad
    \mu,\nu=0,x,y,z.
\end{equation}
Evaluation of the elements of the spin-flip overlap matrix, in the order $(0,x,y,z)$, gives
\begin{equation}
M^{+} = \alpha^\ast\beta^\ast
\vcenter{\hbox{\footnotesize$
\begingroup
\setlength{\arraycolsep}{2pt}
\renewcommand{\arraystretch}{1.2}
\begin{pmatrix}
    0 & \sqrt{2p_0p_x} & -2i\sqrt{p_0p_y} & -\sqrt{2p_0p_z} \\[3pt]
    \sqrt{2p_0p_x} & 2p_x & -i\sqrt{2p_xp_y} & 0 \\[3pt]
    -2i\sqrt{p_0p_y} & -i\sqrt{2p_xp_y} & 0 & -i\sqrt{2p_yp_z} \\[3pt]
    -\sqrt{2p_0p_z} & 0 & -i\sqrt{2p_yp_z} & -2p_z
\end{pmatrix}.
\endgroup
$}}
\end{equation}
Diagonalization of $M^{+}M^{+\dagger}$ then gives
\begin{align}
    \mu_{1,2}
    &=
    |\alpha\beta|
    \left[
    \sqrt{1-(p_0-p_y)^2}
    \pm
    |p_x-p_z|
    \right],
    \nonumber\\
    \mu_3&=\mu_4=0.
\end{align}
As a result, $\mathcal{C}(\rho_{CT})
    =
    \mu_1-\mu_2
    =
    2|\alpha\beta|
    |p_x-p_z|$,
and the upper bound  given in Eq.~\eqref{eq:pauli-H-upper-bound} is saturated. Maximizing over the control state, with the maximum attained for
$|\alpha|=|\beta|=1/\sqrt{2}$, lets us obtain
\begin{equation}
    \mathcal{C}_{\max}
    \left(
    \mathcal{H},\mathcal{P}
    \right)
    =
    |p_x-p_z|,
\end{equation}
which proves Eq.~\eqref{eq:pauli-H-maximum} of the main text.

\section{Switch of amplitude damping and bit-flip channels}
\label{appD}

In this part, we first derive the expressions in Eqs.~\eqref{eq:AD-X-angular-concurrence} and~\eqref{eq:AD-X-Lambda}, and then establish the upper bound in Eq.~\eqref{eq:AD-X-upper-bound}. For compactness, let us introduce
\begin{equation}
    a
    =
    \cos(\vartheta/2),
    \quad
    b
    =
    e^{i\varphi}\sin(\vartheta/2),
    \quad
    r
    =
    \sqrt{1-\gamma},
    \label{eq:AD-X-parameters-appD}
\end{equation}
so that the target state in Eq.~\eqref{eq:AD-X-target} is written as
\begin{equation}
    \ket{\psi}
    =
    a\ket{0}
    +
    b\ket{1}.
\end{equation}
Using the Kraus representation in Eq.~\eqref{eq:AD-kraus}, the two unnormalized vectors in
Eq.~\eqref{eq:AD-X-vectors} take the explicit form
\begin{align}
    \ket{w_0}
    ={}&
    \alpha b\ket{00}
    +
    \alpha r a\ket{01}
    +
    \beta r b\ket{10}
    +
    \beta a\ket{11},
    \nonumber\\
    \ket{w_1}
    ={}&
    \sqrt{\gamma}
    \left(
        \alpha a\ket{00}
        +
        \beta b\ket{11}
    \right).
\end{align}
The spin-flip overlap matrix for this decomposition is
\begin{equation}
    M_{ij}
    =
    \braket{w_i|\widetilde{w}_j},
    \qquad
    i,j=0,1.
\end{equation}
Direct evaluation of its matrix elements gives
\begin{align}
    M_{00}
    &=
    M_{11}
    =
    -2\gamma\alpha^\ast\beta^\ast a^\ast b^\ast,
    \nonumber\\
    M_{01}
    &=
    M_{10}
    =
    -\sqrt{\gamma}\alpha^\ast\beta^\ast
    \left[
        (a^\ast)^2+(b^\ast)^2
    \right].
\end{align}
Therefore,
\begin{equation}
    M
    =
    -\sqrt{\gamma}\alpha^\ast\beta^\ast
    \begin{pmatrix}
        2\sqrt{\gamma}\,a^\ast b^\ast
        &
        (a^\ast)^2+(b^\ast)^2
        \\[3pt]
        (a^\ast)^2+(b^\ast)^2
        &
        2\sqrt{\gamma}\,a^\ast b^\ast
    \end{pmatrix}.
\end{equation}
A direct calculation gives the singular values of $M$ as
\begin{equation}
    \left\{
        \mu_1,\mu_2
    \right\}
    =
    \sqrt{\gamma}|\alpha\beta|
    \left\{
        |u+v|,
        |u-v|
    \right\},
\end{equation}
where $u=a^2+b^2$ and $v=2\sqrt{\gamma}\,ab$, which enables us to obtain the concurrence in the form
\begin{equation}
    \mathcal{C}(\rho_{CT})
    =
    \sqrt{\gamma}|\alpha\beta|
    \left|
        |u+v|
        -
        |u-v|
    \right|.
    \label{eq:AD-X-concurrence-appD}
\end{equation}
Let us next express this result in terms of the target state angles. Using Eq.~\eqref{eq:AD-X-parameters-appD}, one finds
\begin{align}
    u
    &=
    \cos^2(\vartheta/2)
    +
    e^{2i\varphi}\sin^2(\vartheta/2)
    \nonumber\\
    &=
    e^{i\varphi}
    \left(
        \cos\varphi
        -
        i\cos\vartheta\sin\varphi
    \right),
    \nonumber\\
    v
    &=
    \sqrt{\gamma}\,
    e^{i\varphi}\sin\vartheta.
\end{align}
Consequently,
\begin{align}
    |u\pm v|^2
    ={}&
    \left(
        \cos\varphi
        \pm
        \sqrt{\gamma}\sin\vartheta
    \right)^2
+
    \cos^2\vartheta\sin^2\varphi,
\end{align}
and $|u\pm v| =\Lambda_\pm,$ where $\Lambda_\pm$ are given in Eq.~\eqref{eq:AD-X-Lambda}. Substitution into Eq.~\eqref{eq:AD-X-concurrence-appD} proves Eq.~\eqref{eq:AD-X-angular-concurrence} in the main text.

Finally, we optimize the degree of entanglement over arbitrary complex control and target input states. Applying the reverse triangle inequality to Eq.~\eqref{eq:AD-X-concurrence-appD},
\begin{align}
\left|
|u+v|
-
|u-v|
\right|
&\leq
|(u+v)-(u-v)|
\nonumber\\
&=
2|v|
\nonumber\\
&=
2\sqrt{\gamma}\sin\vartheta.
\end{align}
It follows that
\begin{equation}
\mathcal{C}(\rho_{CT})
\leq
2\gamma|\alpha\beta|\sin\vartheta.
\label{eq:AD-X-bound-appD}
\end{equation}
This upper bound is attained for every fixed $\vartheta$. Indeed, when
$\varphi=0$ or $\pi$, the target amplitudes are real and
\begin{equation}
u=1,
\qquad
v
=
\pm\sqrt{\gamma}\sin\vartheta.
\end{equation}
Since
$|v|=\sqrt{\gamma}\sin\vartheta\leq1$, one obtains
\begin{equation}
\left|
|1+v|
-
|1-v|
\right|
=
2|v|.
\end{equation}
The reverse triangle bound is saturated and
\begin{equation}
\mathcal{C}(\rho_{CT})
=
2\gamma|\alpha\beta|\sin\vartheta
\end{equation}
for $\varphi=0$ or $\pi$, proving
Eq.~\eqref{eq:AD-X-real-concurrence}. Hence, for fixed target populations, a complex relative phase cannot give a concurrence larger than that obtained with real amplitudes. The remaining optimization over the control coherence and the target populations follows from
\begin{equation}
2\gamma|\alpha\beta|\sin\vartheta
\leq
\gamma,
\end{equation}
where we used $|\alpha\beta|\leq1/2$ and $\sin\vartheta\leq1$. Equality above is attained for $|\alpha|=|\beta|=1/\sqrt{2}$ and $\vartheta = \pi/2$. Combining these conditions with the phase parameters saturating Eq.~\eqref{eq:AD-X-bound-appD}, a global maximum is attained when
\begin{equation}
\vartheta
=
\frac{\pi}{2},
\qquad
\varphi
=
0
\ \text{or} \
\pi,
\qquad
|\alpha|
=
|\beta|
=
\frac{1}{\sqrt{2}}.
\end{equation}
These choices correspond to a maximally coherent control and either of the real equatorial target states $\ket{\pm}$, and
\begin{equation}
\max_{\ket{\psi_C},\ket{\psi}}
\mathcal{C}(\rho_{CT})
=
\gamma,
\end{equation}
which proves Eq.~\eqref{eq:AD-X-maximum} in the main text.

\section{Switch of generalized amplitude damping channels}
\label{appE}

In this appendix, we derive the composition relations for two generalized amplitude damping channels, establish the concurrence bound in Eq.~\eqref{eq:GAD-concurrence-bound}, and evaluate the family of channels and inputs that saturates this bound. For an arbitrary qubit density operator $\rho$, the Kraus representation in Eq.~\eqref{eq:GAD-kraus} gives the output terms as
\begin{align}
    \left[
        \mathcal{G}_{\gamma,p}(\rho)
    \right]_{00}
    &=
    (1-\gamma)\rho_{00}
    +
    p\gamma\,\mathrm{Tr}\,(\rho),
    \nonumber\\
    \left[
        \mathcal{G}_{\gamma,p}(\rho)
    \right]_{11}
    &=
    (1-\gamma)\rho_{11}
    +
    (1-p)\gamma\,\mathrm{Tr}\,(\rho),
    \nonumber\\
    \left[
        \mathcal{G}_{\gamma,p}(\rho)
    \right]_{01}
    &=
    \sqrt{1-\gamma}\,\rho_{01},
    \quad
    \left[
        \mathcal{G}_{\gamma,p}(\rho)
    \right]_{10}
    =
    \sqrt{1-\gamma}\,\rho_{10}.
    \label{eq:GAD-action-appE}
\end{align}
We first consider two channels with the same stationary population, $p_1=p_2=p$. Defining
\begin{equation}
    \gamma_{12}
    =
    \gamma_1+\gamma_2-\gamma_1\gamma_2,
    \quad
    1-\gamma_{12}
    =
    (1-\gamma_1)(1-\gamma_2),
\end{equation}
successive application of Eq.~\eqref{eq:GAD-action-appE} gives
\begin{align}
    \mathcal{G}_{\gamma_2,p}
    \circ
    \mathcal{G}_{\gamma_1,p}
    =
    \mathcal{G}_{\gamma_{12},p}
    =
    \mathcal{G}_{\gamma_1,p}
    \circ
    \mathcal{G}_{\gamma_2,p},
\end{align}
which proves Eq.~\eqref{eq:GAD-common-p-composition}. For arbitrary $p_1$ and $p_2$, subtracting the two  compositions obtained from Eq.~\eqref{eq:GAD-action-appE} yields
\begin{align}
    &
    \left[
        \mathcal{G}_{\gamma_2,p_2}
        \circ
        \mathcal{G}_{\gamma_1,p_1}
        -
        \mathcal{G}_{\gamma_1,p_1}
        \circ
        \mathcal{G}_{\gamma_2,p_2}
    \right](\rho)
    \nonumber\\
    &\qquad=
    \gamma_1\gamma_2
    (p_2-p_1)
    \mathrm{Tr}\,(\rho)\,
    \sigma_z.
\end{align}
We can see that the two channels  commute under composition if and only if $\gamma_1\gamma_2(p_1-p_2)=0$, proving Eq.~\eqref{eq:GAD-commutation-condition}.

We next derive the concurrence bound. Let us consider the pure control and target inputs
\begin{equation}
    \ket{\psi_C}
    =
    \alpha\ket{0}+\beta\ket{1},
    \qquad
    \ket{\psi}
    =
    a\ket{0}+b\ket{1},
\end{equation}
with
$|\alpha|^2+|\beta|^2=|a|^2+|b|^2=1$.
Taking the channels as
$\mathcal{E}=\mathcal{G}_{\gamma_1,p_1}$ and
$\mathcal{F}=\mathcal{G}_{\gamma_2,p_2}$, the switched output becomes
\begin{equation}
    \rho_{CT}
    =
    \sum_{i,j=0}^{3}
    \ket{w_{ij}}\bra{w_{ij}},
\end{equation}
where $\ket{w_{ij}}
    =
    \alpha\ket{0}\otimes
    G_j^{(2)}G_i^{(1)}\ket{\psi}
    +
    \beta\ket{1}\otimes
    G_i^{(1)}G_j^{(2)}\ket{\psi}.$
In the Kraus representation of Eq.~\eqref{eq:GAD-kraus}, the operators $G_0^{(k)}$ and $G_2^{(k)}$ are diagonal, whereas $G_1^{(k)}$ and $G_3^{(k)}$ represent downward and upward jumps, respectively. If both selected operators are diagonal, their ordered products coincide. If one is diagonal and the other is a jump operator, the two ordered products are proportional to the same rank-one jump operator. In both cases, the corresponding switch vectors are product vectors. Also, two downward jumps or two upward jumps vanish. Thus, the only branches that can be entangled contain one downward and one upward jump. Direct multiplication gives
\begin{align}
    \ket{w_{13}}
    &=
    \sqrt{
        \gamma_1\gamma_2p_1(1-p_2)
    }\,
    \ket{\xi},
    \nonumber\\
    \ket{w_{31}}
    &=
    \sqrt{
        \gamma_1\gamma_2(1-p_1)p_2
    }\,
    \ket{\zeta},
\end{align}
where
\begin{align}
    \ket{\xi}
    &=
    \alpha b\ket{01}
    +
    \beta a\ket{10},
    \nonumber\\
    \ket{\zeta}
    &=
    \alpha a\ket{00}
    +
    \beta b\ket{11}.
    \label{eq:GAD-xi-zeta-appE}
\end{align}
Denoting the sum of all remaining branch contributions by $\rho_0$, the complete output can be written as
\begin{align}
    \rho_{CT}
    ={}&
    \rho_0
    +
    \gamma_1\gamma_2p_1(1-p_2)
    \ket{\xi}\bra{\xi}
    \nonumber\\
    +&
    \gamma_1\gamma_2(1-p_1)p_2
    \ket{\zeta}\bra{\zeta},
\end{align}
where $\rho_0$ is a positive, generally unnormalized, separable operator. The equally weighted combination of the last two projectors is also separable. Indeed, defining
\begin{equation}
    \ket{\Gamma_\pm}
    =
    \ket{\zeta}
    \pm
    \ket{\xi},
\end{equation}
one obtains
    $\ket{\Gamma_\pm}
    =
    \left(
        \alpha\ket{0}
        \pm
        \beta\ket{1}
    \right)
    \otimes
    \left(
        a\ket{0}
        \pm
        b\ket{1}
    \right).$
Thus, both $\ket{\Gamma_+}$ and $\ket{\Gamma_-}$ are product vectors, and
\begin{equation}
    \ket{\xi}\bra{\xi}
    +
    \ket{\zeta}\bra{\zeta}
    =
    \frac{1}{2}
    \sum_{\nu=\pm}
    \ket{\Gamma_\nu}\bra{\Gamma_\nu}
\end{equation}
is separable. Since
    $p_1(1-p_2)
    -
    (1-p_1)p_2
    =
    p_1-p_2,$
the common weight of the two opposite jump terms can be included in the separable part. For $p_1\geq p_2$, this gives
\begin{equation}
    \rho_{CT}
    =
    \rho_{\mathrm{sep}}
    +
    \gamma_1\gamma_2
    (p_1-p_2)
    \ket{\xi}\bra{\xi},
\end{equation}
whereas for $p_2\geq p_1$,
\begin{equation}
    \rho_{CT}
    =
    \rho_{\mathrm{sep}}
    +
    \gamma_1\gamma_2
    (p_2-p_1)
    \ket{\zeta}\bra{\zeta}.
\end{equation}
Here, $\rho_{\mathrm{sep}}$ is a positive, generally unnormalized, separable operator, which can be different in the two cases. Let $\ket{\Omega}=\ket{\xi}$ when $p_1\geq p_2$ and
$\ket{\Omega}=\ket{\zeta}$ otherwise. Normalizing the two positive terms before applying the convexity of concurrence gives
\begin{equation}
    \mathcal{C}(\rho_{CT})
    \leq
    \gamma_1\gamma_2
    |p_1-p_2|
|\!\braket{\Omega|\widetilde{\Omega}}\!|.
    \label{eq:GAD-convexity-appE}
\end{equation}
Indeed, the weight of the normalized pure state contribution contains the factor $\braket{\Omega|\Omega}$, while its concurrence contains the inverse factor, so that the norm cancels in Eq.~\eqref{eq:GAD-convexity-appE}. For the two vectors in Eq.~\eqref{eq:GAD-xi-zeta-appE},
\begin{equation}
    |\!\braket{\xi|\widetilde{\xi}}\!|
    =
    |\!\braket{\zeta|\widetilde{\zeta}}\!|
    =
    2|\alpha\beta ab|.
\end{equation}
It follows that
\begin{equation}
    \mathcal{C}(\rho_{CT})
    \leq
    2\gamma_1\gamma_2
    |p_1-p_2|
    |\alpha\beta ab|.
\end{equation}
Finally, using
$|\alpha\beta|\leq1/2$ and $|ab|\leq1/2$, we obtain
\begin{equation}
    \mathcal{C}(\rho_{CT})
    \leq
    \frac{\gamma_1\gamma_2}{2}
    |p_1-p_2|.
\end{equation}
Since the right-hand side of the bound is independent of the input states, taking the maximum over pure control and target inputs on the left yields Eq.~\eqref{eq:GAD-concurrence-bound}.

Finally, we evaluate the concurrence for the particular family considered in the main text. Let
\begin{equation}
    \gamma_1=\gamma_2=1,
    \qquad
    p_1=1,
    \qquad
    p_2=p,
\end{equation}
and prepare both the control and target inputs in $\ket{+}$. In the computational basis, the switched output is
\begin{equation}
    \rho_{CT}
    =
    \frac{1}{4}
    \begin{pmatrix}
        2p & 0 & p & 0\\
        0 & 2(1-p) & 1-p & 0\\
        p & 1-p & 2 & 0\\
        0 & 0 & 0 & 0
    \end{pmatrix}.
\end{equation}
The nonzero singular values entering the concurrence are
\begin{equation}
    \mu_{1,2}
    =
    \frac{\sqrt{1-p}}{4}
    \left(
        2
        \pm
        \sqrt{1-p}
    \right),
    \quad
    \mu_3=\mu_4=0.
\end{equation}
Therefore,
\begin{equation}
    \mathcal{C}(\rho_{CT})
    =
    \mu_1-\mu_2
    =
    (1-p)/2,
\end{equation}
which proves Eq.~\eqref{eq:GAD-saturating-family}. For these channel parameters, the upper bound in
Eq.~\eqref{eq:GAD-concurrence-bound} is also equal to $(1-p)/2$ and is saturated for every $p\in[0,1]$. Since
\begin{equation}
    \frac{\gamma_1\gamma_2}{2}
    |p_1-p_2|
    \leq
    \frac{1}{2},
\end{equation}
setting $p=0$ gives $\mathcal{C}(\rho_{CT})=1/2$, which establishes the global maximum. Interchanging the two channels gives the equivalent case $(p_1,p_2)=(0,1)$. When $\ket{0}$ is the ground state and the stationary populations are restricted to Gibbs states at nonnegative temperature, then $1/2\leq p_k\leq1$ and hence $|p_1-p_2|\leq1/2$. The bound then
gives
$\mathcal{C}_{\max}
    \leq
    1/4$.
This value is also attained by the same family with $p_1=1$ and $p_2=1/2$, yielding $\mathcal{C}_{\max}=1/4$.

\section{Concurrence for the time-flip of unitary channels}
\label{appF}

Here, we derive the spectral and geometric expressions for the maximal concurrence in
Eqs.~\eqref{eq:unitary-time-flip-spectral-maximum} and~\eqref{eq:unitary-time-flip-maximum}. As
$Y=U^\ast U$ is unitary and $\det Y=(\det U)^\ast\det U=1$, we have $Y\in SU(2)$, and its eigenvalues can be written as in Eq.~\eqref{eq:time-flip-spectrum}. Let $\ket{\xi_\pm}$ denote the corresponding orthonormal eigenstates. An arbitrary normalized target state can be parametrized in this eigenbasis as
\begin{equation}
    \ket{\psi}
    =
    \cos(\vartheta/2)\ket{\xi_+}
    +
    e^{i\varphi}\sin(\vartheta/2)\ket{\xi_-},
\end{equation}
with $ 0\leq\vartheta\leq\pi$ and $ 0\leq\varphi<2\pi$. The overlap in Eq.~\eqref{eq:time-flip-overlap} then becomes
\begin{equation}
    \begin{aligned}
        \omega
        &=
        \cos^2(\vartheta/2)e^{i\theta_Y/2}
        +
        \sin^2(\vartheta/2)e^{-i\theta_Y/2}
        \\
        &=
        \cos(\theta_Y/2)
        +
        i\cos\vartheta\sin(\theta_Y/2).
    \end{aligned}
\end{equation}
Consequently,
\begin{equation}
    \begin{aligned}
        |\omega|^2
        &=
        \cos^2(\theta_Y/2)
        +
        \cos^2\vartheta\sin^2(\theta_Y/2)
        \\
        &=
        1
        -
        \sin^2\vartheta\sin^2(\theta_Y/2).
    \end{aligned}
\end{equation}
For $Y\neq\pm\mathbb{I}$, the magnitude of the overlap is minimized at $\vartheta=\pi/2$. Thus, the optimal target states are
\begin{equation}
    \ket{\psi_{\text{max}}}
    =
    \frac{
        \ket{\xi_+}
        +
        e^{i\varphi}\ket{\xi_-}
    }{\sqrt{2}},
    \qquad
    0\leq\varphi<2\pi,
\end{equation}
where the relative phase is arbitrary. If $Y=\pm\mathbb{I}$, every target state instead gives $|\omega|=1$ and hence zero concurrence. In all cases,
\begin{equation}
    \min_{\ket{\psi}}|\omega|
    =
    \left|\cos(\theta_Y/2)\right|.
\end{equation}
The control contribution satisfies $2|\alpha\beta|\leq1$, with equality for a balanced control state, $|\alpha|=|\beta|=1/\sqrt{2}$. Substitution into Eq.~\eqref{eq:unitary-time-flip-concurrence} gives
\begin{equation}
        \mathcal{C}_{\max}(U)
        =
        \left|\sin(\theta_Y/2)\right|
        =
        \sqrt{
            1-(1/4)
            \bigl(\mathrm{Tr}\,Y\bigr)^2
        },
    \label{eq:time-flip-spectral-maximum-appF}
\end{equation}
where we used $\mathrm{Tr}\,Y=2\cos(\theta_Y/2)$. We next derive the geometric form of this result. The global phase of $U$ cancels in $Y$, so we may indeed use the $SU(2)$ representation in
Eq.~\eqref{eq:time-flip-axisangle}. In the computational basis fixed for the quantum time-flip,
$\sigma_x^\ast=\sigma_x^{ T}=\sigma_x$,
$\sigma_y^\ast=\sigma_y^{ T}=-\sigma_y$, and
$\sigma_z^\ast=\sigma_z^{ T}=\sigma_z$. It follows that
\begin{equation}
    U^\ast
    =
    \cos(\theta_U/2)\,\mathbb{I}
    +
    i\sin(\theta_U/2)
    \left(
        n_x\sigma_x
        -
        n_y\sigma_y
        +
        n_z\sigma_z
    \right).
\end{equation}
Using $\mathrm{Tr}\,(\sigma_j)=0$ and $\mathrm{Tr}\,(\sigma_j\sigma_k)=2\delta_{jk}$, where $\delta_{jk}$ is the Kronecker delta and $j,k\in\{x,y,z\}$, we obtain
\begin{equation}
    \begin{aligned}
        \frac{\mathrm{Tr}\,Y}{2}
        &=
        \cos^2(\theta_U/2)
        +
        \sin^2(\theta_U/2)
        \left(
            n_x^2-n_y^2+n_z^2
        \right)
        \\
        &=
        1
        -
        2n_y^2\sin^2(\theta_U/2),
    \end{aligned}
\end{equation}
where we have exploited the fact that $n_x^2+n_y^2+n_z^2=1$. Introducing $\kappa=n_y^2\sin^2(\theta_U/2)$, with $0\leq\kappa\leq1$, we thus have $\mathrm{Tr}\,Y/2=1-2\kappa$. Equation \eqref{eq:time-flip-spectral-maximum-appF} then yields
\begin{align}
        \mathcal{C}_{\max}(U)
        =
        \sqrt{1-(1-2\kappa)^2}
        =
        2\sqrt{\kappa(1-\kappa)},
\end{align}
which proves Eq.~\eqref{eq:unitary-time-flip-maximum} quoted in the main text.

\bibliography{bibliography}

\end{document}